\documentclass[acmlarge]{acmart}
\usepackage{subfigure}
\AtBeginDocument{%
  \providecommand\BibTeX{{%
    \normalfont B\kern-0.5em{\scshape i\kern-0.25em b}\kern-0.8em\TeX}}}

\setcopyright{acmcopyright}
\copyrightyear{2025}
\acmYear{2025}
\acmDOI{XXXXXXX.XXXXXXX}

\acmJournal{POMACS}
\acmVolume{37}
\acmNumber{4}
\acmArticle{111}
\acmMonth{8}

\usepackage{caption}

\usepackage[normalem]{ulem}
\newcommand \change[1]{{\textcolor{black}{#1}}}

\begin{document}

\title{TextOnly: A Unified Function Portal for Text-Related Functions on Smartphones
}


\author{Minghao Tu}
\affiliation{%
  \institution{Tsinghua University}
  \city{Beijing}
  \country{China}
}

\author{Chun Yu}
\affiliation{%
  \institution{Tsinghua University}
  \city{Beijing}
  \country{China}
}

\author{Xiyuan Shen}
\affiliation{%
  \institution{University of Washington}
  \city{Seattle}
  \country{United States}
}
\affiliation{%
  \institution{Tsinghua University}
  \city{Beijing}
  \country{China}
}

\author{Zhi Zheng}
\affiliation{%
  \institution{Tsinghua University}
  \city{Beijing}
  \country{China}
}

\author{Li Chen}
\affiliation{%
  \institution{Tsinghua University}
  \city{Beijing}
  \country{China}
}

\author{Yuanchun Shi}
\affiliation{%
  \institution{Tsinghua University}
  \city{Beijing}
  \country{China}
}

\renewcommand{\shortauthors}{Tu, et al.}

\begin{abstract}

  Text boxes \change{serve as portals to diverse functionalities in today's smartphone applications.}
  However, \change{when it comes to specific functionalities, }users always need to navigate through multiple steps to access \change{particular} text boxes for input. 
  \change{We propose} TextOnly, a unified function portal that enables users to access text-related functions \change{from various applications} by simply inputting text \change{into a sole text box}. 
  For instance, entering a restaurant name could \change{trigger} a Google Maps search, \change{while a greeting could} initiate a conversation in WhatsApp. Despite their \change{brevity, TextOnly maximizes the utilization of these raw text inputs, which contain rich information, to interpret user intentions effectively.}
  \change{TextOnly integrates large language models (LLM) and a BERT model. The LLM consistently provides general knowledge, while the BERT model can continuously learn user-specific preferences and enable quicker predictions.}
  Real-world user studies demonstrated TextOnly's \change{effectiveness} with a top-1 accuracy of 71.35\%, and \change{its ability to continuously improve both its accuracy and inference speed.}
  \change{Participants perceived TextOnly as having satisfactory usability and expressed a preference for TextOnly over manual executions.} Compared with voice assistants, TextOnly supports a greater range of text-related functions and allows for more concise inputs.
  
\end{abstract}



\received{20 February 2007}
\received[revised]{12 March 2009}
\received[accepted]{5 June 2009}

\maketitle

\begin{figure}
  \centering
  \includegraphics[width=1.0\linewidth]{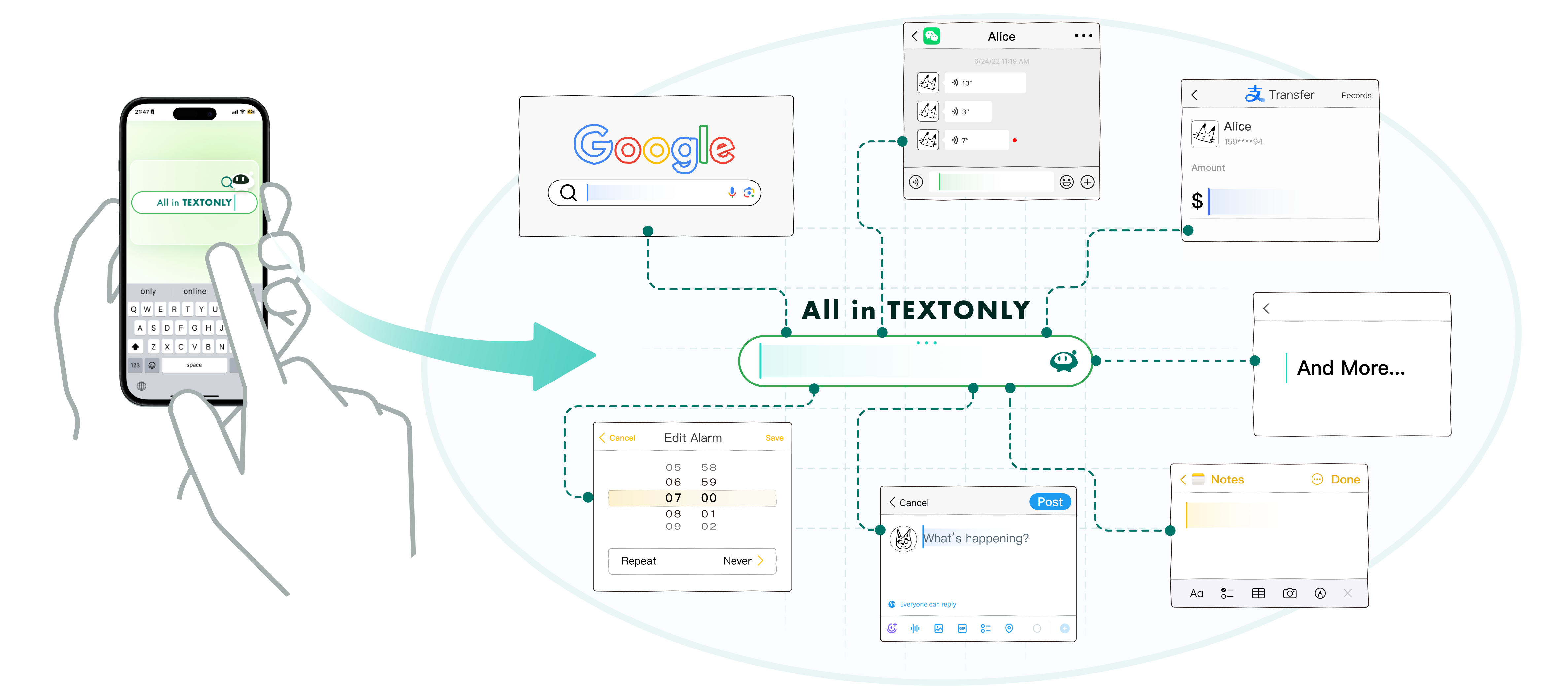}
  \caption{\change{TextOnly redirects users' input to their intended text boxes. The input of TextOnly is exactly what the user types into his intended text box.}}
  \label{fig:textonly}
\end{figure}

\section{INTRODUCTION}

\change{With the expansion of mobile services in various aspects of daily life, the number of smartphone applications} has significantly increased, resulting in over 3.4 million apps on the Google Play Store\cite{googleplay}. 
Consequently, a significant amount of time is spent on selecting and navigating across different apps when users \change{seek} to access a specific function or service. 
To address this challenge, researchers have proposed various solutions to simplify the workflow of \change{function navigation}, such as app selection via text query\cite{aliannejadi2018target,aliannejadi2021context} and app recommendation that predicts the next apps to be used\cite{liao2012mining,huang2012predicting,lu2014mining}. 
While these solutions help users access their \change{desired} apps, \change{they fall short in seamlessly guiding users to their intended destinations within the apps.}


\change{Among the array of functions available on smartphones, text-related functions constitute a significant portion, referring to functions that require users to input text during its execution. We observe that users' concise text inputs encompass rich information that can be leveraged to predict their intentions. Consequently, by inferring users' intentions from their unprocessed text inputs, we can greatly streamline the process of accessing text-related functions.}

We propose TextOnly, a unified function portal on smartphones that enables users to access text-related functions from various applications by inputting ``raw text input'' into a unified text box. 
As shown in Figure~\ref{fig:textonly}, ``raw text input'' can be the keywords to be searched on Google, the messages to be sent on WeChat, or the content to be posted on Twitter, etc. 
In this \change{context}, TextOnly \change{significantly} reduces the \change{necessary} operations and demands shorter inputs \change{compared to traditional} Natural Language Interfaces (NLIs), which \change{usually require additional words for a comprehensive description of an intent.}

\change{However, precise function prediction is extremely difficult due to the ambiguity of natural language, the diversity of various functions, and the significant differences in personalized preferences among users.} \change{Thus, we propose an inference
framework that combines a pre-trained large language model and a BERT model\cite{devlin2018bert}.
On one hand, we effectively tackle the cold start problem on new users and new functions, where the model struggles to operate efficiently for new users without usage histories and new functions without historical data\cite{chen2021predicting}.
In contrast to prior research on next-use app prediction which leverages other users' usage histories or app semantic representations to address the user cold start problem\cite{baeza2015predicting, chen2021predicting}, we integrate a pre-trained LLM to acquire general knowledge and inference abilities, empowering TextOnly to handle new users and new functions. 
On the other hand, for user personalization, we employ a BERT model that learns from user data and LLM's predictions during usage. The BERT model assists in prompt construction for LLM by selecting data samples similar to users' new input, and generates alternative predictions faster than LLM.}

To evaluate TextOnly's performance \change{and usability}, we conducted a real-world user study with 16 participants over a week-long period. TextOnly reached an accuracy of 71.35\% for top-1 recommendations and 89.94\% for top-5 recommendations.
\change{For usability, participants are satisfied with the accuracy and inference time of TextOnly, and have a preference for TextOnly over manual executions. They also expressed enthusiasm for its continued usage, and were patient and confident in TextOnly even when it was inaccurate. We also noticed that TextOnly can change participants' input habits to some extent.} \change{In a laboratory user study, we compared TextOnly with existing built-in text portals and voice assistants. Results show that TextOnly supports a wider range of text-related functions and its inputs are significantly shorter than that of voice assistants.}


In summary, our main contributions are as follows:
\begin{itemize}
    \item We first propose the approach that utilizes users' raw text inputs to infer users' intentions at function level. This introduces a novel interaction method for the use of text-related functions on mobile devices.
    \item We collected real data on users' usage of text-related functions and analyzed their association with users' corresponding inputs and contextual data.
    \item We implemented a unified function portal for text-related functions and verified its practicality and capability for continuous improvements in accuracy and inference speed through a real-world user study.
\end{itemize}

\section{RELATED WORK}

Our work focuses on quick access to functions and services on smartphones, and user intention inference through text input. \change{In our framework, we introduce large language models as the main inference module.} In this section, we revisit related work on service retrieval, natural language interfaces, \change{and large language models.}

\subsection{Service Retrieval}

The goal of service retrieval is to efficiently and effectively retrieve relevant services that meet the user's needs or preferences. Service retrieval \change{was} initially introduced in Web, where researchers have explored various approaches for \change{service matching}, such as syntactic matching\cite{klusch2006automated,meditskos2009structural,ma2008efficiently}, ontology-based matching\cite{pathak2005framework,broens2004context}, semantic matching\cite{toch2007semantic,klusch2008semantic}, etc. With the advancement of \change{the} Internet of Things (IoT), service retrieval techniques have also been applied in IoT environments\cite{cassar2012hybrid,zhao2015topic,cao2019qos}.

On mobile devices, however, due to the lack of unified \change{entry points} for various services across different applications, there are only a few industrial systems that provide efficient access to in-app services and functions. For instance, Apple Spotlight\cite{applespotlight} \change{stands out as} the most popular example of such systems on iOS devices. Another example is Sesame Shortcuts\cite{sesameshorcuts}, an Android app that offers quick access to installed apps via keyword-based queries. Despite the existence of these systems, there \change{remains a dearth} of research focusing on service retrieval on smartphones.

On the other hand, research has been done on mobile information retrieval, \change{encompassing investigations into } users' search patterns and behavior on smartphones\cite{kamvar2006large,guy2016searching,shokouhi2014mobile,church2008large}. Researchers have integrated information \change{from} various sources and modalities to perform information retrieval at \change{the} application level. For example, Li et al.\cite{li2017mining} utilized online reviews to perform app recommendations on \change{the} app market. Park et al.\cite{park2016mobile} leveraged users' status on social media for app recommendations. Aliannejadi et al.\cite{aliannejadi2018target,aliannejadi2021context} proposed \change{a} unified mobile search, which provides app recommendations based on user queries using contextual information of previous app usage. Sensor contextual information like location and time is also used for app usage predictions\cite{chen2019cap,yu2018smartphone, wang2021app2vec, xia2020deepapp}.

In contrast, our work achieves service retrieval on smartphones through a \change{unified} text portal. Although its supported services are constrained to be text-related, TextOnly \change{supports various applications and in-app operations using Robotic Process Automation (RPA)\cite{hofmann2020robotic}}. Furthermore, we use the raw text input to infer \change{users'} intentions, which is a novel methodology for mobile information retrieval.

\subsection{Natural Language Interfaces}

Natural language interfaces (NLIs) allow users to interact with a computer system or application using natural language\cite{thompson1987natural}. With the development of natural language processing, NLIs have gained the capability to handle diverse functions and users' natural expressions \cite{su2017building, berant2013semantic, zettlemoyer2012learning}. NLIs have been applied in a wide range of tasks such as question answering \cite{yih2015semantic}, command execution \cite{li2020mapping, gur2018learning, humphreys2022data}, workflow generation \cite{brachman2022goal}, and information management \cite{zhou2012mobile, zhou2007natural}. However, the functions supported by NLIs often require manual configuration by developers \cite{lau2010conversational, ravindranath2012code}, and the ability to understand diverse user expressions demands the collection of numerous training data \cite{azaria2016instructable, su2017building}. Consequently, these constraints limit the number of functions supported by NLIs, especially those provided by third-party sources.

As a special type of NLI, voice assistants have become increasingly popular, particularly on smartphones\cite{pearl2016designing, pan2022automatically}. Nevertheless, voice interaction \change{faces} a lot of challenges, including difficulties in editing\cite{shneiderman2000limits}, recognition errors, and insufficient feedback\cite{myers2018patterns, luger2016like}. Additionally, voice interaction is inconvenient in public or noisy environments, therefore necessitating the demand for convenient text-based NLIs.

Our proposed approach, TextOnly, can be regarded as a unique NLI for text-related functions that takes shorter inputs rather than complete instructions. In terms of execution, TextOnly completely carry out \change{users'} intended functions, instead of just navigating to the applications. This approach effectively simplifies users' interactions and makes their input process more effortless. Besides, it is worth noting that TextOnly is not confined to typing as its only input modality, but also accommodates voice input.

\subsection{Large Language Models}

Large Language Models (LLMs) typically refer to generative pre-trained language models that process and generate text sequences based on the provided input. Notable LLMs such as GPT-3\cite{brown2020language}, PaLM\cite{chowdhery2022palm}, and ChatGPT\cite{chatgpt}, have demonstrated the capability to perform various tasks without fine-tuning, empowering users to guide them solely through prompts. LLMs not only possess extensive general and specialized knowledge, but have also emerged with reasoning abilities that can be further enhanced through Chain-of-Thought methodologies\cite{wei2022chain, kojima2022large}.

LLMs open up new possibilities for human-computer interaction\cite{bommasani2021opportunities}. They can effectively handle the diverse natural language expressions by end-users, enabling developers to effortlessly create applications with \change{powerful capabilities of natural language interaction\cite{wu2023visual}}. \change{Researchers also utilize its comprehensive general knowledge to assist in context understanding and logic reasoning\cite{khaokaew2024maple,chen2023gap}.} However, the substantial amount of computation required by LLMs results in non-negligible response latency and cost\cite{chen2023frugalgpt}, highlighting the need for better strategies in utilizing LLMs.

In this paper, we employ LLM for its comprehensive knowledge and ability to infer the intended functions of users, thereby addressing the cold start problem. To reduce the time consumption and computational cost of LLM, we incorporate a BERT model that learns from user data and LLM's predictions to assist in prompt construction and enhance system's overall inference speed. 

\section{TEXTONLY}

In this section, we introduce TextOnly, a unified function portal for text-related functions that helps users achieve their text-related intentions quickly and effortlessly. We first introduce its user interface and functionalities, followed by an overview of its overall framework. Then we provide further details on our utilization of LLM, the design of the BERT-based encoder, and our approach \change{to} personalized learning, respectively.

\subsection{User Interface}

\begin{figure}
  \centering
  \subfigure[App Triggering]{
  \begin{minipage}[b]{0.3\linewidth}
    \centering
    \includegraphics[width=\linewidth]{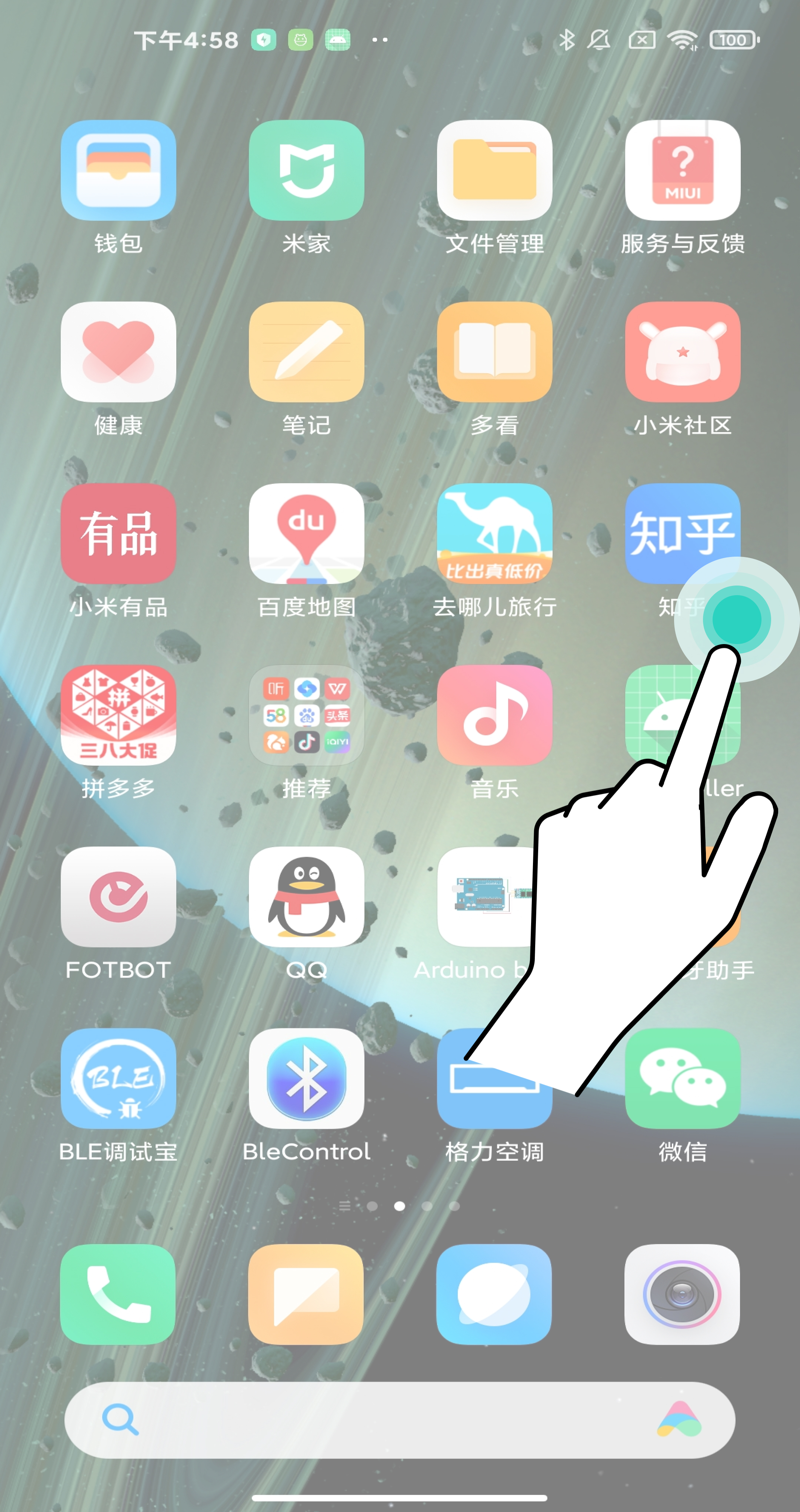}
    \label{fig:interface1}
  \end{minipage}
  }
  \subfigure[Text Input]{
  \begin{minipage}[b]{0.3\linewidth}
    \centering
    \includegraphics[width=\linewidth]{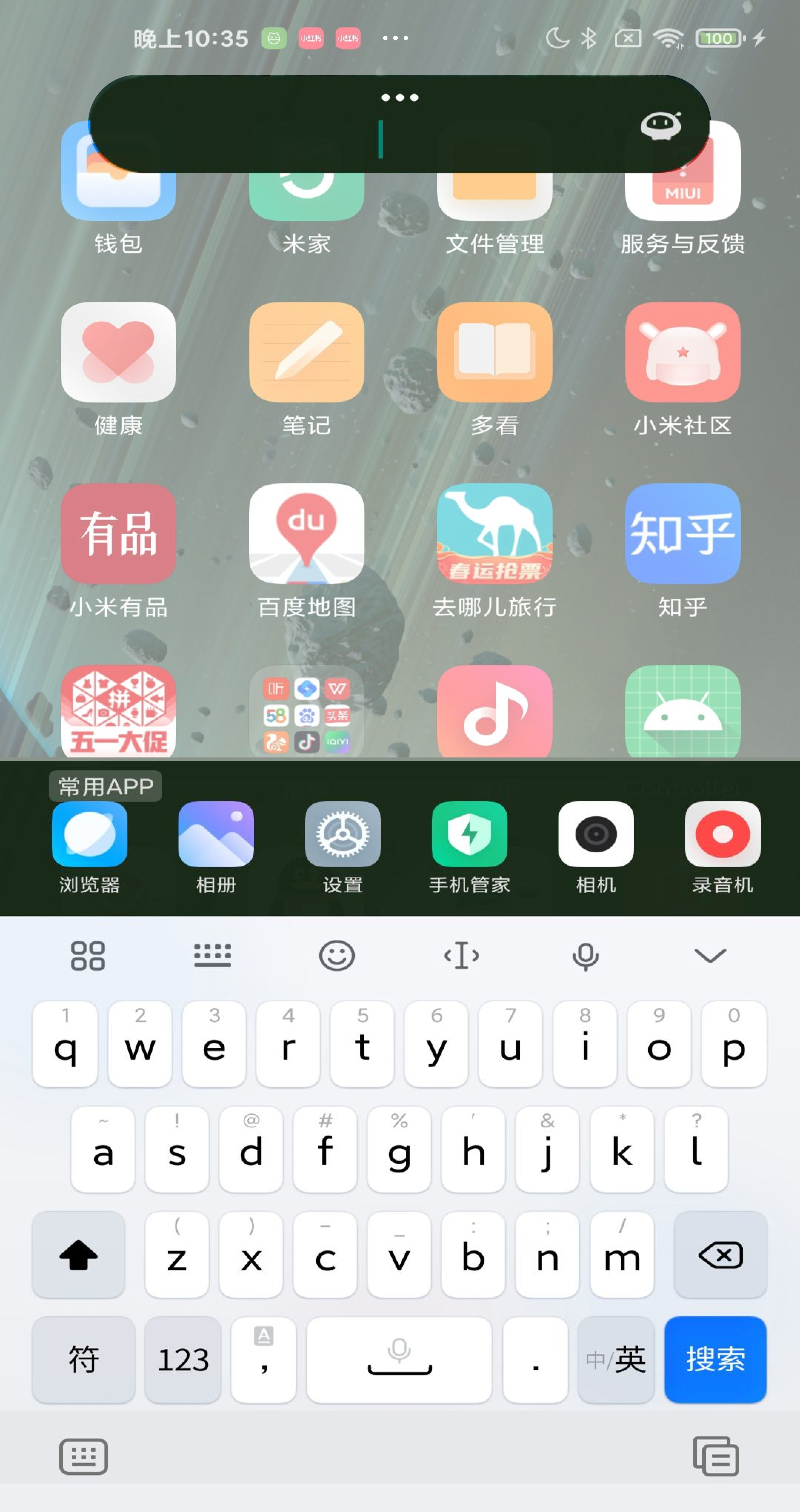}
    \label{fig:interface2}
  \end{minipage}
  }
  \subfigure[Function Selection]{
  \begin{minipage}[b]{0.3\linewidth}
    \centering
    \includegraphics[width=\linewidth]{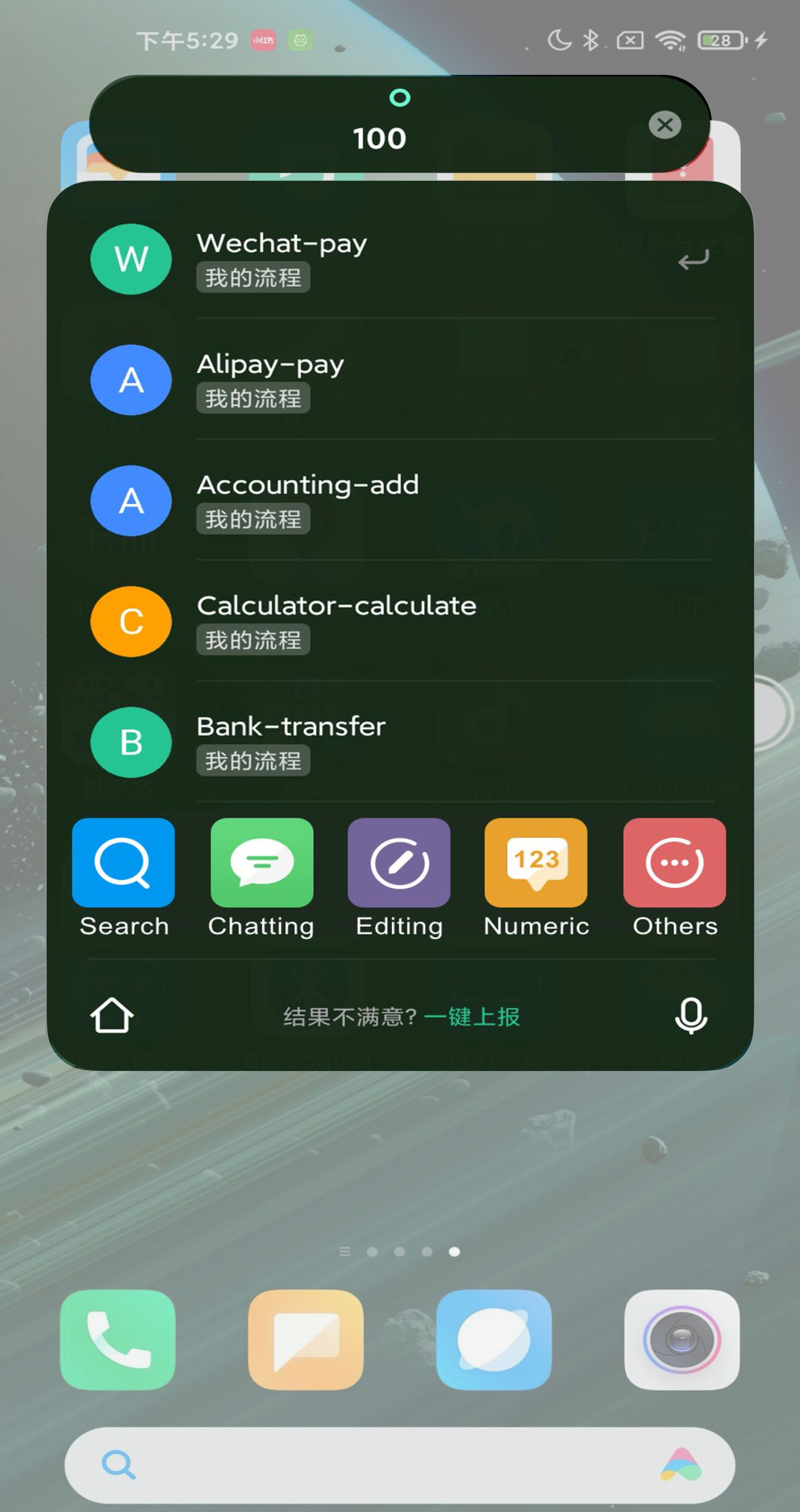}
    \label{fig:interface3}
  \end{minipage}
  }
  \caption{\change{Interfaces of TextOnly. Figure (a) illustrates the initiation of TextOnly from the home screen; Figure (b) is the input interface of TextOnly; Figure (c) is the function selection interface of TextOnly, where the user's input is ``100''.}}
  \label{fig:interface}
\end{figure}

The main interfaces of TextOnly are depicted in Figure~\ref{fig:interface}. \change{As a unified function portal, TextOnly's user interfaces are quite simple, mainly composed of a text box that can be triggered conveniently. As shown in Figure~\ref{fig:interface1}, users can trigger TextOnly by clicking on the floating ball on the screen. TextOnly can also be triggered by the user's tap-tap gesture, which is supported by \emph{future scanner}\cite{scanner}, an Android app designed by our team. Upon triggering TextOnly, users can type their inputs in the interface of Figure~\ref{fig:interface2}, and then TextOnly utilizes the input text to predict users' intended text-related functions. The predictions are displayed to the users in a list, as depicted in Figure~\ref{fig:interface3}.}  Users can scroll through the list to view more candidate functions. Once the intended function is recommended accurately, the user only needs to click on it, \change{and TextOnly will execute the selected function automatically. The automatic execution of functions is facilitated by \emph{future scanner}\cite{scanner}, using RPAs.}
 
In this context, \change{the} user's input in TextOnly is exactly \change{what he types in his manual execution}. This differentiates TextOnly from voice assistants and other NLIs, where users' inputs typically consist of commands that describe their intents completely. In cases where TextOnly's recommendations are not accurate enough, users can opt to specify their desired applications or functions explicitly by inputting a symbol ``*'' followed with additional content. The additional content is used to match with the applications or descriptions of the candidate functions.

For each user, TextOnly offers a default collection of candidate functions, comprising commonly used text-related functions on smartphones. During the initial usage, TextOnly verifies whether the application associated with each default function is installed on the user's device, thus ensuring an executable collection tailored to the user. Users are also granted the capability to manage their functions manually. Users can remove seldom-used or unnecessary functions from the list, as well as add new ones by recording corresponding RPAs.

Due to privacy concerns, we excluded chat-oriented intentions during data collection. However, in the implementation of TextOnly, we still offer support for these chat-oriented intentions. For such intentions, we need to further specify the contact to chat with. Therefore, we represent chatting functions in the form of ``app-contact''. Once a chatting function is selected, TextOnly will automatically send the message to the chosen contact in the chosen app.

\subsection{Model Framework\label{framework}}

\begin{figure}
  \centering
  \includegraphics[width=1.0\linewidth]{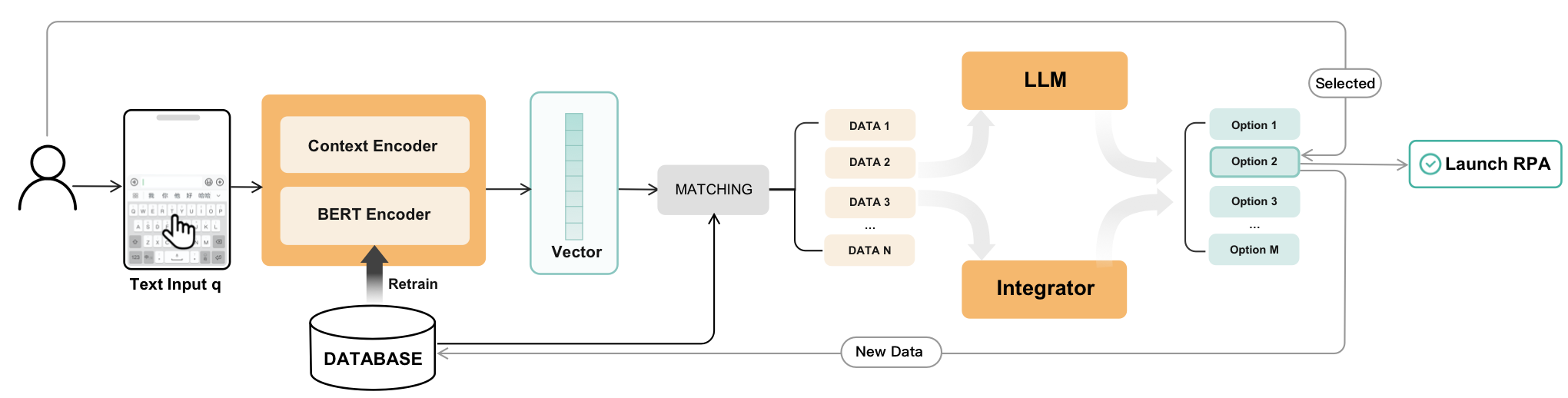}
  \caption{Model framework of TextOnly}
  \Description{}
  \label{fig:framework}
\end{figure}

The core function of TextOnly is to predict \change{the} user's intended text-related function based on \change{the} user's input and additional contextual information. To achieve this, our model necessitates comprehensive general knowledge to \change{understand} the features and meanings of text inputs and text-related functions, as well as user-personalized knowledge to characterize users' preferences on application usages and text expressions. Hence, our inference framework is founded on a pre-trained large language model enriched with extensive general knowledge, and it incorporates a BERT-based encoder for assimilating users' personalized knowledge. 

Given a user input $q$, TextOnly first converts it into a one-dimensional vector $\mathbf{v}$: 
\begin{equation}
    \mathbf{v} = \phi(q) \oplus \mathbf{v}_{app\_usage} \oplus \mathbf{v}_{time}
\end{equation}

\begin{equation}
    y = Sigmoid(\mathcal{W}\cdot\phi(q)+b)
\end{equation}
where $\phi(q)$ denotes the vector derived from text via the BERT encoder, $\mathbf{v}_{app\_usage}$ and $\mathbf{v}_{time}$ denote the vectors derived from the contextual information of app usage and time respectively, and $\oplus$ denotes the concatenation of the vectors.

Subsequently, a linear layer is applied to the vector to determine whether the input corresponds to a chat-oriented intention, as this will determine the set of candidate functions for the subsequent inference task. We then search the database $H$ for history inputs that are similar to $q$. The similarity between $q$ and $q_{i}$ is calculated as follows: 
\begin{equation}
 sim(\mathbf{v},\mathbf{v}_{i}) = max\{cos\_sim(\mathbf{v}, \mathbf{v}_{i})\cdot\alpha_{user\_weight}, 1\}
 \label{equation:similarity}
\end{equation}
where $\alpha_{user\_weight}$ is a coefficient that equals constant $C_{user\_weight}$ when $q$ and $q_{i}$ are from a same user, or 1 otherwise.

Afterward, we select \change{the} top $K$ instances of data in $H$ that share \change{the} highest similarity with input $q$, and feed them to LLM and a local integrator for predictions, respectively. For the prediction of LLM, the instances are used to construct prompts that contain relevant information with input $q$. As for the local integrator, it integrates the labels of the $K$ instances to a new prediction at a speed significantly faster than LLM. The prediction of the local integrator is generated as follows:
\begin{equation}
    Prediction(q) = \frac{\sum\limits_{i=1}\limits^{K} Label_{k_{i}}\cdot sim(q,q_{k_{i}})}{\sum\limits_{i=1}\limits^{K} sim(q,q_{k_{i}})}
    \label{equation:prediction}
\end{equation}
where $Prediction$ and $Label_{k_{i}}$ are both $1*N$ vectors ($N$ equals the number of candidate functions), with each dimension representing the probability of the corresponding candidate function.

We employ a confidence score to select the final prediction between the predictions of LLM and the local integrator. The confidence score $v_{confidence}$ is defined as a weighted average of the similarities between input $q$ and the selected $K$ instances:
\begin{equation}
    v_{confidence} = \frac{\sum\limits_{i=1}\limits^{K} sim(q,q_{k_{i}}) \cdot (K-i+1)}{\sum\limits_{i=1}\limits^{K} i}
\end{equation}
If $v_{confidence}$ is greater than a predefined threshold $T$, the prediction of the local integrator is taken as TextOnly's output directly. If not, the LLM will be queried for a potentially better but slower prediction.

Our aim \change{in} combining LLM with a BERT model is to leverage LLM's general knowledge and ability to ensure the model's performance with limited data, which takes place during the initial usage of each user. With continued use over time, the BERT model will learn from the accumulated user data and the inference results of the LLM, to enhance its capacity for representing users' inputs. As the BERT model improves, TextOnly is expected to rely less on the LLM and provide faster predictions.

\subsection{LLM and Prompt Constructing}

The LLM serves as the principal inference component in our framework. In our implementation, the GPT-3.5 model\change{\cite{gpt3.5}} from OpenAI is employed through HTTP API calls. Throughout its usage, the model remains unaltered and unfine-tuned. \change{Instead}, we craft specific prompts to \change{augment} it with enhanced inference capabilities. Owing to constraints in LLM's inference speed, an approach of a single-step query is adopted. Nevertheless, alternative multi-step or heuristic querying strategies, such as the Chain-of-Thought method\cite{wei2022chain}, could enhance the LLM's performance, \change{though} at the expense of increased time consumption.

For prediction tasks other than chat-oriented intentions, we construct the prompts with four parts as follows: 
\begin{itemize}
    \item Task description: A concise explanation of the prediction task
    \item Candidate options: A list of all candidate functions, each with a concise introduction (optional).
    \item Few-shot examples: $M$ most similar instances of historical data \change{from} database $H$, each formatted to include \change{user} input, app usage, time, and output.
    \item Input query:  User's text input along with a specification that the output should rank the top five candidates in order of likelihood, for ease of constructing $Label_{k}$ in Equation~\ref{equation:prediction}.
\end{itemize}
After experiments and testing, we find that removing the description of candidates in the part of candidate options leads to better results, and the value of $M$ is set to 20.

For contact selection of chat-oriented intentions, the prompts also start with a task description and candidate options, i.e. a list of contacts, followed by chat histories with each contact and the input query. To prevent the prompts from becoming excessively lengthy, we set an upper limit for the length of the chat history with a single contact, based on the size of the contact list. \change{Exact prompts can be found in Appendix~\ref{appendix:prompt}.}

\subsection{BERT-based Encoder}

We employed a BERT-based encoder to learn from user data and LLM's predictions through usage, \change{to feature} effective representations for text inputs in text-related functions. This, in turn, can be used to construct better few-shot prompts for LLM and generate predictions with low latency and economic cost. Subsequently, we delve into the details of the model training process.

Given that the predictions of the local integrator are based on similarity, as is shown in Equation~\ref{equation:prediction}, we naturally consider employing the Info Noise-contrastive Estimation (Info NCE) loss\change{\cite{oord2018representation}} for model training. However, this loss function leads to slow training speed, which conflicts with our needs for ongoing updates to the BERT model during its use. To this end, we have adopted an alternative loss function that enables faster model training. This loss function is formulated in a direct manner, utilizing a parameter matrix $\mathcal{W}:\mathbb{R}^{d}\rightarrow\mathbb{R}^{N}$ that maps the input vector $\phi(q)$ to the probability of each candidate, where $d$ represents the dimension of $\phi(q)$ and $N$ denotes the number of candidate options. We calculate a cross-entropy loss as follows:
\begin{gather}
    \hat{y} = \mathcal{W}\cdot\phi(q)+b \label{equation:Wq}\\
    \mathcal{L}_{CrossEntropy} = -\sum\limits_{i=1}\limits^{n} y_{i}\log\hat{y_{i}}
\end{gather}

Employing this function, we train additional parameters $\mathcal{W}$ that can be directly applied in prediction. With this prediction approach, however, \change{the impact of new data only becomes apparent after subsequent training cycles, and is fairly marginal}. \change{We conducted comparisons between the combinations of the two loss functions and the two prediction methods, and results revealed no significant differences.} Consequently, we \change{choose} the original prediction methodology of the local integrator, which more effectively utilizes new data and offers better interpretability. 

After experiments, the threshold $T$ is set to be 0.95, while $K$ is fixed at 5 and $C_{user\_weight}$ at 1.05. 

\subsection{Personalized Learning}

Due to the significant differences among users in habits, preferences, and application usage, we maintain a personal database for each user and train individual models based on their personal databases.

For initial usage, TextOnly provides a default collection of functions for each user, \change{which consists of the top 20 most frequently used functions in data collection}. Users can manage their own collection of functions \change{in TextOnly}. For a new user without any personal historical data, we pre-train his/her BERT model with a subset of data selected from the overall database, which comprises his/her initial personal database. We select the initial data according to the initial set of user's text-related functions, with the following two rules:

\begin{itemize}
    \item The amount of data selected for each action correlates with the number of the user's text-related functions of the specific action, scaled by a coefficient $\alpha$.
    \item The amount of data selected for each function correlates with its frequency in the overall database.
\end{itemize}

The first rule defines the total number of functions for each action, while the second rule specifies the distribution of specific functions within each action. Specifically, $\alpha$ is set to 10, so the amount of data used for pre-training equals 10 times the number of the user's text-related functions, which is typically a few hundreds. For rare text-related functions with insufficient data in the overall database, we utilize the LLM to generate \change{synthetic} data for pre-training.

During usage, user data from each interaction is recorded in \change{the} user's personal database, and promptly impacts the predictions of the BERT model through Equation~\ref{equation:prediction}. As \change{the} user's feedback only consists of one selected function, we combine the user's selection with the LLM's output to formulate $Label$, attributing weights to the top five options. After testing, we have adopted a specific weight distribution of $[0.8, 0.07, 0.06, 0.04, 0.03]$. The BERT model is continuously updated through retraining on a daily basis.

\section{TEXT-RELATED FUNCTIONS AND USER INPUT}

To investigate \change{the frequency of users' use of text-related function, as well as the relationship between users' input and their intended functions}, we conducted a week-long data collection and analyzed the data. \change{This study proved both the necessity and feasibility of TextOnly, which provides users with quick access to text-related functions via their raw text inputs. In the following,} we provide details on how we collected data and present our findings. 

\subsection{Data Collection\label{datacollection}}
In order to collect user data during real-world usage, we implemented an Android application, which monitors \change{users' keyboard input events} through Android's accessibility service\cite{accessibility} and manages a floating ball that always floats on the top of the screen. Whenever the user inputs text into a text box in any application, our app considers it a potential \change{piece of data} and turns the floating ball red, reminding the user to upload data. By clicking the floating ball, users can navigate to the data submission interface of our app and submit their data. Also, users are free to ignore the reminder when occupied or busy and postpone the submission until a more convenient time.

During data submission, our app automatically captures the text previously entered by users, along with the application in which it was entered. The participant only needs to use one or two words to briefly describe the desired action corresponding to the input, such as \emph{search, record, share, etc.}, making the data submission process quite convenient. In this case, we define user intentions in the form of ``application-action'', like ``Google-search'', ``Memo-record'', and ``Facebook-chat''. Due to the inconsistency in the words used by different participants to describe their intentions, we manually \change{checked} the data collected to ensure that the same intentions are described using uniform terms. In addition to the text inputs and intentions, our data collection app also collects some contextual data during usage, including time, recent usage of apps, and recent notifications of apps. When a participant uploads a data entry, the mentioned data will also be recorded.

\subsubsection*{\change{Participants}}

We recruited a total of 22 participants (8 males and 14 females), \change{comprising students and staff from campus, aged between 19 and 30 years. All participants use smartphones extensively in their daily lives. Detailed information about the participants is listed in Appendix~\ref{appendix:participants1}.} We conducted a week-long data collection in real-world scenarios and carried out brief interviews with each participant afterwards. \change{Participants received compensation according to the volume of data uploaded (1 CNY per data entry) and the time of the interviews conducted (100 CNY per hour).}

\change{Before the study, we obtained IRB approval and each participant signed a consent form.} Due to participants' feedback and privacy concerns, we excluded chat-oriented intentions from the scope of data collection. We provided participants with the autonomy to selectively upload data, allowing them to withhold any information they deem sensitive or confidential for privacy or security reasons. We provided a clear explanation of how we were going to use the collected data as well as what was collected from their devices. Participants could withdraw from the study at any time, and we would delete all of their data from the servers. It is important to note that our background service did not collect any other data from participants’ smartphones. 

\subsection{Findings}

As a result, we gathered a total of 5,136 data entries and 275 distinct intentions. We analyzed the data from various perspectives and unearthed several insightful findings. These findings not only validate the practicality of TextOnly, but also offer valuable guidance for the implementation of our system.

\begin{figure}
  \centering
  \includegraphics[width=0.8\linewidth]{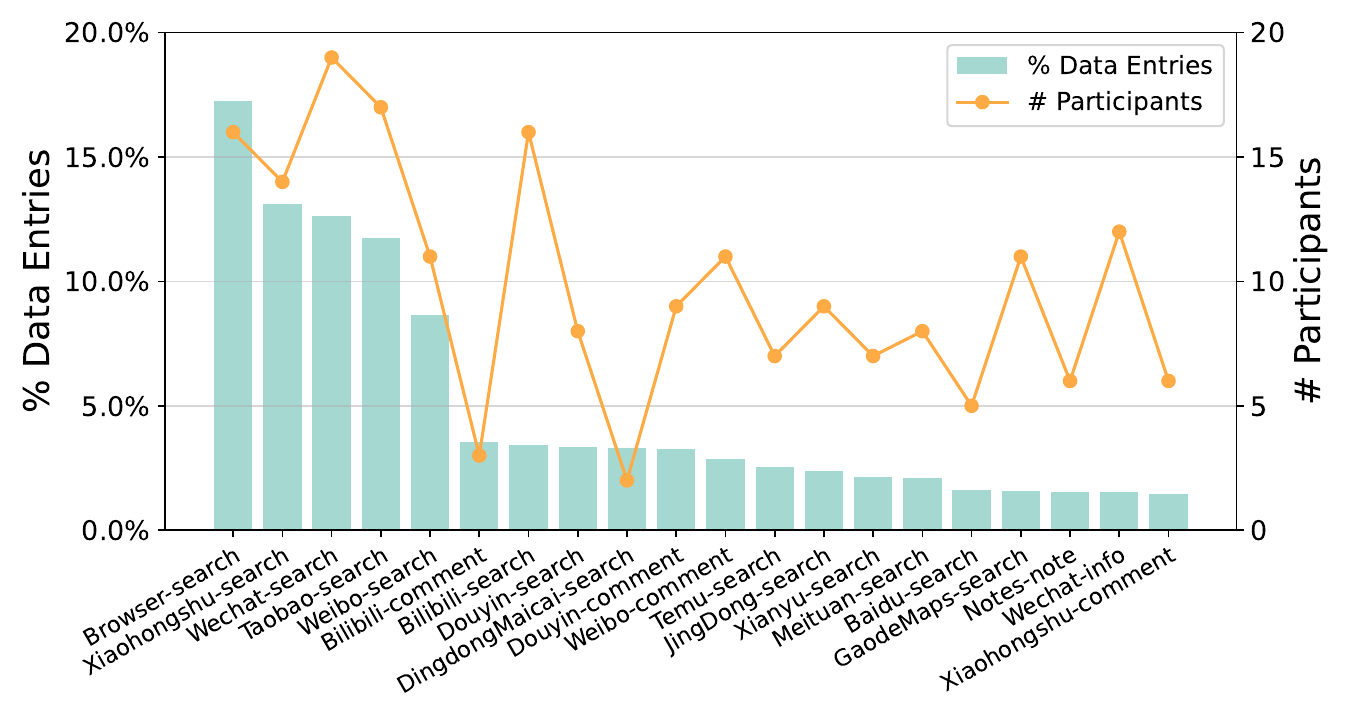}
  \caption{Number of data entries for top 20 text-related functions. \change{``\# Participants'' refers to the number of participants that used the function.}}
  \Description{}
  \label{fig:functions}
\end{figure}

\subsubsection{Overall usage}

\change{\textbf{Participants use text-related functions frequently in their daily activities.}} After seven days of data collection, we accumulated 5,136 data entries. This averages to 233.5 entries per participant, or 33.4 entries per participant per day. This indicates a high frequency of text-related functions in participants' daily use of smartphones. Additionally, in the post-collection interviews, many participants reported a high frequency of chat-oriented intentions that were excluded during data collection, \change{further suggesting the high frequency of text-related functions' usage}.

\change{\textbf{A wide range of different text-related functions are used in participants' daily activities.}} In terms of intentions, we collected 275 distinct intentions and each participant's data comprised 26.7 unique intentions on average. Under our definition, each unique intention corresponds to \change{exactly one} unique text-related function. This underscores the rich diversity of text-related functions utilized in \change{participants'} daily smartphone use. 

In summary, it is suggested that people frequently utilize a wide variety of text-related functions during their daily \change{smartphone use}. It emphasizes the \change{fundamental} role that text input plays in users' interactions with smartphones, through which users access and utilize a wide range of functionalities. \textbf{This highlights the significance and importance of a unified function portal for \change{quick access to text-related functions}}.

\subsubsection{Distribution over text-related functions}

\begin{figure}
  \centering
  \includegraphics[width=0.8\linewidth]{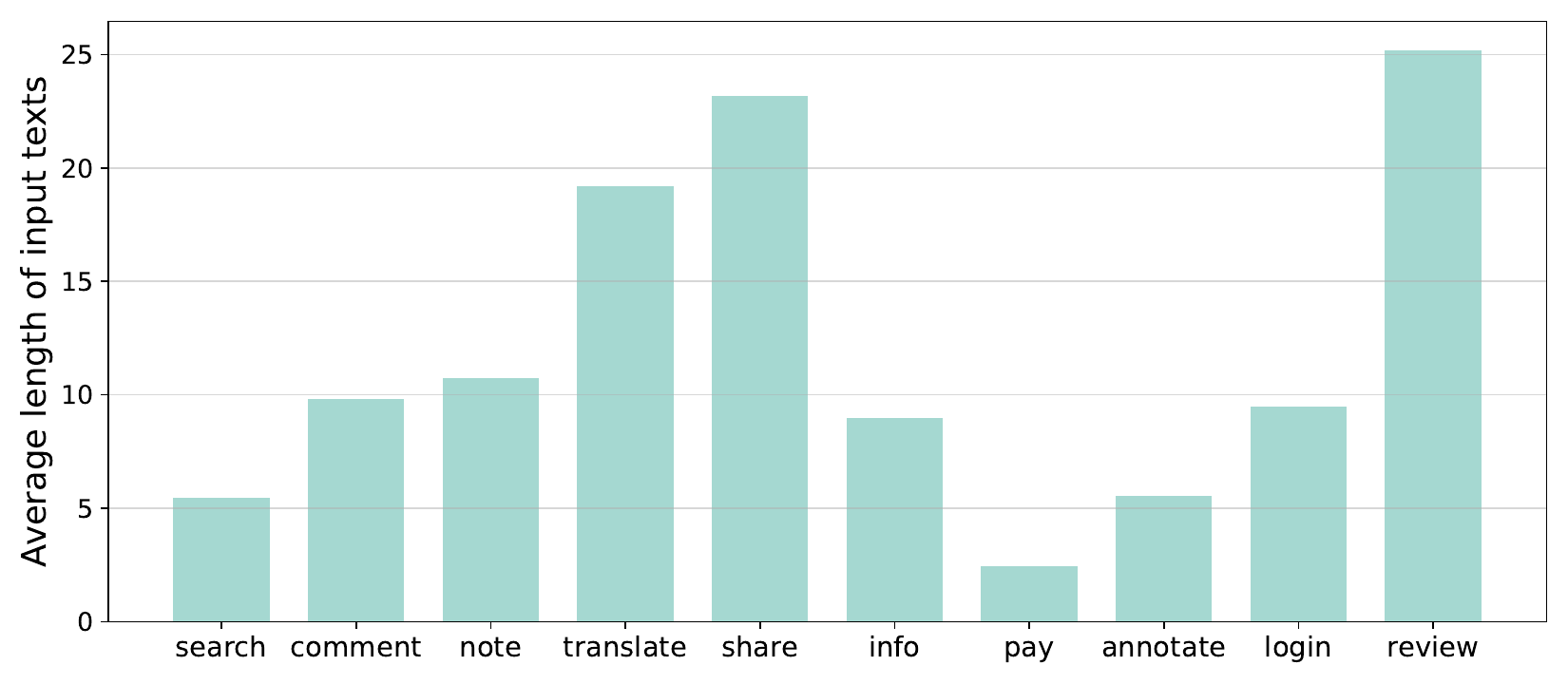}
  \caption{Average length of input texts of each action}
  \Description{}
  \label{fig:action}
\end{figure}

\change{\textbf{Searching is the most common intention.}} Figure~\ref{fig:functions} illustrates the top 20 most frequent text-related functions, represented \change{as} ``app-action''. These top 20 functions account for 76.35\% of \change{all} collected data, indicating a power-law distribution for the frequency of text-related functions. The function ``Browser-search'' \change{is the most commonly used, indicating that participants often rely on the default search engine for search intentions}. Most of the top 20 functions (14 out of the 20) are search functions, with a considerable frequency each. This suggests that searching is a major intention when people input text on their smartphones. 

\change{\textbf{The text input can be quite different for functions of different apps.}} Unlike the generality of the default search engine in browsers, many in-app searches have distinct characteristics, like Gaode Maps for locations and Dingdong Maicai for food. These features distinguish various search functions, enabling us to \change{associate the inputs with} specific text-related functions. Similarly, the same \change{goes for} actions other than search, \change{where} the specific application exerts a significant influence on users' text inputs. \change{For example, users' reviews on IMDB and Amazon can be quite different. This indicates the feasibility of predicting users' intended app according to their inputs.}

\change{\textbf{The text input can be quite different for functions of different actions.}} Figure~\ref{fig:action} illustrates the average length of text inputs for the top 10 most frequent actions. A considerable disparity can be observed \change{across different actions}. Notably, the \change{inputs} of action \emph{review} have the longest average length, reaching 25.20 \change{words}, \change{while the inputs of action \emph{pay} are the shortest, with only 2.47 words}. This difference is consistent with the nature of these actions. Action \emph{review} typically involves participants writing reviews on platforms of food delivery or shopping sites, normally necessitating more extended text. Conversely, action \emph{pay} corresponds to participants entering \change{a number} for mobile payments, which typically \change{comprises 2 or 3 digits}. Besides, action \emph{translate} and \emph{share} are associated with longer inputs, whereas action \emph{search} and \emph{annotate} tend to involve shorter inputs. These patterns align with our common knowledge and habits. Such findings \change{reveal} that the \change{users' text inputs} exhibit distinct characteristics for different actions, reflective of \change{its use scenarios}.

\subsubsection{Distribution over participants}

\begin{figure}
  \centering
  \subfigure[]{
  \begin{minipage}[b]{0.45\linewidth}
    \centering
    \includegraphics[width=\linewidth]{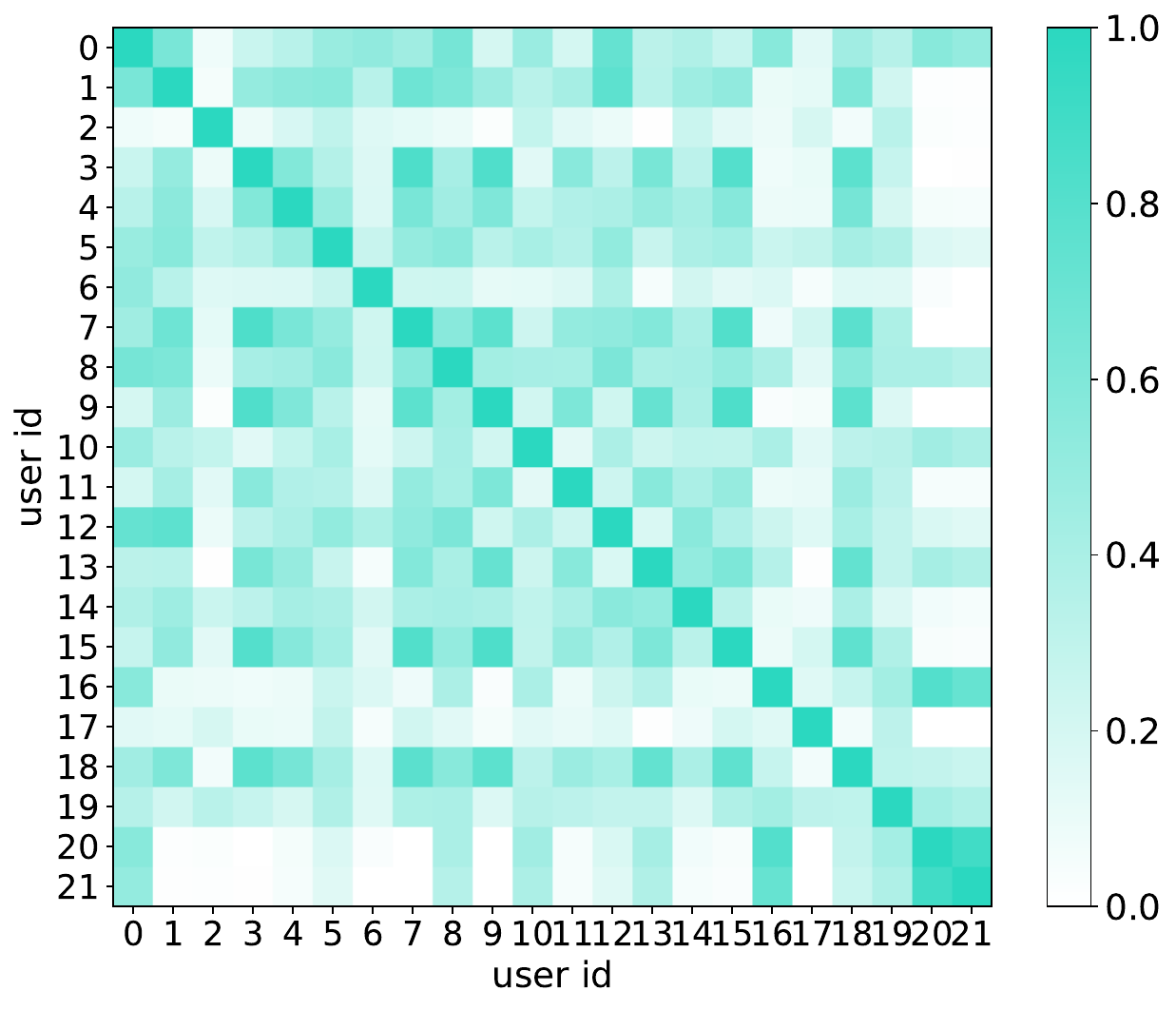}
    \label{fig:heatmap1}
  \end{minipage}
  }
  \subfigure[]{
  \begin{minipage}[b]{0.45\linewidth}
    \centering
    \includegraphics[width=\linewidth]{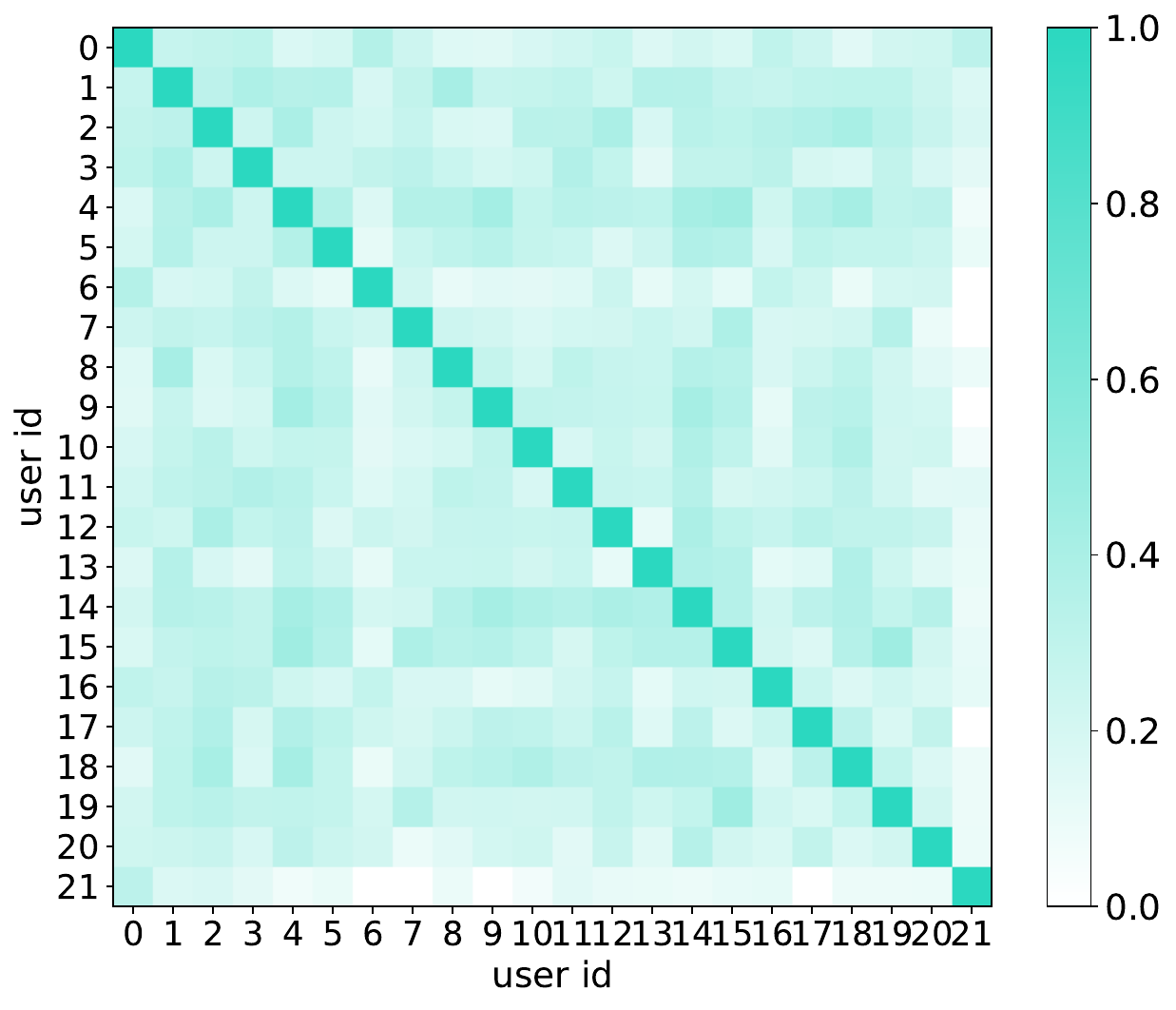}
    \label{fig:heatmap2}
  \end{minipage}
  }
  \caption{Similarity heatmaps of text-related functions between participants}
  \label{fig:similarity}
\end{figure}

\change{\textbf{The frequency and range of participants' text-related function use varies significantly.}} Among the \change{22} participants, an average of 233.5±184.3 data entries involving 26.7±18.4 text-related functions were collected, demonstrating a \change{large} variance. While this variability can be attributed to the participants' engagement and their choices of data submission, it \change{still} reveals substantial differences among participants in both the frequency and \change{range} of text-related functions. We converted \change{the usage of text-related functions for each participant} into an embedding, \change{with its length} equals the total number of \change{collected functions}. We then generated the similarity heatmaps shown in Figure~\ref{fig:similarity}. \change{For Figure~\ref{fig:heatmap1}, participants' embeddings are based on} the frequency of their text-related function usage. \change{For Figure~\ref{fig:heatmap2}, participants' embeddings are based on} whether the function was used by the participants, in an ont-hot format.

Both heatmaps \change{are} quite sparse, indicating a low degree of similarity among participants in their usage of text-related functions. Figure~\ref{fig:heatmap2} exhibits lower similarity compared to Figure~\ref{fig:heatmap1}, suggesting that although the participants have a low overlap in the \change{range of used text-related functions}, there are some shared patterns in their frequencies for certain widely-used functions. This highlights the critical need for user personlization in \change{the implementation of TextOnly}, to accommodate the varied preferences and requirements of users.

\subsubsection{App usage and notifications}

\begin{table}
  \caption{Usage and notification history of users desired apps. ``\% in x min'' denotes the proportion of the desired apps that have been used or issued notifications within the past x minutes. ``\# in x min'' denotes the number of times that the desired apps was used or issued notifications within the past x minutes.}
  \label{tab:context}
  \begin{tabular}{lrrrr}
    \toprule
    App List & \% in 1 min & \% in 5 min & \% in 10 min & \# in 10 min \\
    \midrule
    Used Apps & 33.62\% & 54.98\% & 62.80\% & 2.31 \\
    Notification Apps & 5.99\% & 7.13\% & 7.80\% & 0.28 \\
  \bottomrule
\end{tabular}
\end{table}

\change{\textbf{App usage history can be helpful in predicting users' intentions.}} In addition to user inputs and intentions, we collected pre-input contextual data, including app usage and notifications. We \change{assume} that this contextual information \change{can help predict} users' subsequent use of text-related functions. Table~\ref{tab:context} shows the association between users' desired app for their inputs and the aforementioned contextual information. It \change{suggests} that participants are likely to have accessed the app before they \change{type} in it. \change{For 33.62\% of the data}, the app was used within the last minute before \change{users'} input. We also noticed a tendency for users to switch between \change{multiple} apps before eventually entering text in \change{one of them}. This is reflected \change{by} the data of the last column, which shows that the app for input might have been accessed multiple times recently. Notably, we have excluded instances that \change{the participants stayed} in one single app and used text-related functions within it. This \change{implies} that \change{previous} app usage can be \change{useful in predicting which text-related functions} the user is likely to use next.

Regarding app notifications, we did not find a strong correlation between them and the apps where users \change{input text} subsequently. In post-collection interviews, most \change{participants} reported a substantial connection between notifications of chatting apps and subsequent text inputs. However, as chat-oriented intentions, such data was not included in the data collection. Additionally, some \change{participants} mentioned that they \change{disabled} most notifications on their smartphones, which further diminishes the association between their text inputs and app notifications.

\section{EVALUATION}

To comprehensively evaluate the performance of TextOnly, we conducted studies in both real-world and laboratory environments. \change{In the real-world user study, we evaluated TextOnly's practical performance and learning capabilities during usage. We also conducted interviews with users to gather their subjective impressions of TextOnly. In the laboratory user study, we assessed TextOnly's performance on chat-oriented intentions, and compared it with existing text portals and voice assistants.}

In the following, we detail the design and results of these two studies, and then engage in a discussion grounded in the evaluation results.

\subsection{\change{Real-world} User Study Design}

\subsubsection*{\change{Participants}}

We conducted a real-world user study over a one-week period, to assess the practicality of TextOnly. We recruited 16 participants \change{from campus}, comprising 6 males and 10 females, aged between 19 and 28 years old. \change{All participants used Android smartphones. Detailed information is listed in Appendix~\ref{appendix:participants2}.} \change{Participants received compensation in the same manner as Section~\ref{datacollection}.}

\change{Before the study, we obtained IRB approval for this study and participants signed consent forms.} We provided detailed information to the participants about the data that TextOnly would collect from their devices and how we would utilize it. We also demonstrated how to add or delete text-related functions within TextOnly. Participants were encouraged to customize their collection of text-related functions before the study and were free to modify it at any time during the study.

\subsubsection*{Procedure}

\change{We implemented TextOnly as an Android app and assisted participants with the installation and permission settings prior to the user study.} During the user study, participants were required to use TextOnly for inputting and accessing their desired text-related functions. In addition, they were asked to rate the predictions of TextOnly on a scale from 1 to 5 for each use, with 5 indicating complete satisfaction and 1 indicating significant dissatisfaction. Due to privacy concerns, we still excluded chat-oriented intentions in this study and tested them in a controlled environment instead, which is demonstrated in \change{Section~\ref{labstudy}}. \change{After the user study, we conducted interviews with the participants.}

\change{Considering that participants might forget to use TextOnly in some instances, we retained the reminder of the floating ball in the data collection process, allowing participants to test TextOnly even after they had completed their intentions. While this method did not contribute to quicker realization of participants' intentions, it remained an effective method for testing TextOnly's predictive capabilities.}

\subsubsection*{Evaluation metrics}

Three standard evaluation metrics are used to measure the accuracy of predictions: the accuracy of the top 1 recommendation (Hit@1), the accuracy of the top-5 recommendations (Hit@5) and Mean Reciprocal Rank (MRR). One subjective evaluation metric, satisfaction score, is used to assess users' personal perceptions. Additionally, we measure the inference time of the models, \change{which refers to how long it takes for the model to make predictions.}

When TextOnly's first recommendation is correct, users can accomplish their intentions effortlessly by simply clicking the confirmation button. When the correct function appears within the top five recommendations, users don't have to scroll for more options since TextOnly presents exactly five options in its recommendation list. This rationale informs our choice of using Hit@1 and Hit@5 as evaluation metrics. As for MRR, it is one of the most commonly used metrics for evaluating recommendation systems that reflects systems' accuracy adequately.

\subsubsection*{Compared methods}

We recorded participants' use cases during the user study to test other models for comparisons. We implemented several baselines for comparison. We also generated variants by removing specific modules to assess their impact on TextOnly's performance. The compared methods are as follows:
\begin{itemize}
    \item Most Frequently Used (MFU): Rank the candidates according to their frequency of use as predictions.
    \item Most Recently Used (MRU): Rank the candidates according to their latest time of use as predictions.
    \item Bayesian: Rank the candidiates using a Naive Bayes method based on users' application usage.
    \item TextOnly-general: A modified TextOnly with a general model for all users instead of personalized models. The parameter of threshold $T$ is set to be 0.97.
    \item TextOnly-nocontext: A modified TextOnly with no utilization of contextual information.
    \item BERT-only: An individual BERT model. The integrated results are used as predictions.
    \item LLM-only: An individual LLM. Prompts are constructed with randomly selected samples in the database.
\end{itemize}

\subsection{\change{Real-world} User Study Results}

\begin{table}
  \caption{Performance comparison between TextOnly and compared methods. Values in parentheses denote standard deviation. \change{The superscript * and ** indicates statistically significant results between TextOnly and this model with p < 0.05/0.01 according to t-test, respectively. The hypothesis test was conducted across participants (n=16).}}
  \label{tab:comparison}
  \begin{tabular}{lcccc}
    \toprule
    Model & Hit@1 & Hit@5 & MRR & Time /s \\
    \midrule
    MFU & 0.3077\(^{**}\) (0.1629) & 0.7010\(^{**}\) (0.1414) & 0.4732\(^{**}\) (0.1436) & - \\
    MRU & 0.3909\(^{**}\) (0.1446) & 0.6579\(^{**}\) (0.1145) & 0.5045\(^{**}\) (0.1208) & - \\
    Bayesian & 0.3519\(^{**}\) (0.1412) & 0.7249\(^{**}\) (0.1435) & 0.5195\(^{**}\) (0.1275) & - \\
    \hline
    TextOnly-general & 0.5363\(^{**}\) (0.1694) & 0.7699\(^{*}\) (0.1148) & 0.6261\(^{**}\) (0.1503) & 2.47 (2.36) \\
    TextOnly-nocontext & 0.6011\(^*\) (0.1420) & 0.8375 (0.0944) & 0.6947 (0.1194) & 1.64 (2.21) \\
    LLM-only & 0.5503\(^{**}\) (0.1646) & 0.8294 (0.1109) & 0.6600\(^{*}\) (0.1451) & 4.62\(^{**}\) (0.65)\\
    BERT-only & 0.5402\(^{**}\) (0.1785) & 0.7731\(^{*}\) (0.1328) & 0.6316\(^{**}\) (0.1594) &  \textbf{0.04 (0.02)}\\
    \hline
    TextOnly & \textbf{0.6961 (0.0887)} & \textbf{0.8660 (0.0864)} & \textbf{0.7650 (0.0872)} & 1.90 (2.13)\\
  \bottomrule
\end{tabular}
\end{table}

\subsubsection{Overall results}

Following the week-long user study, we recorded a total of 3,847 trials from 16 participants, with an average of 240.44 (\emph{SD} = 164.85) entries per participant. The participants utilized 243 different text-related functions via TextOnly, with an average of 26.88 (\emph{SD} = 13.29) functions per participant. For all trials, TextOnly achieved a Hit@1 accuracy of 0.7135, a Hit@5 accuracy of 0.8994, an MRR of 0.7897, and an average satisfaction score of 4.36.  

\change{On} average across participants, the Hit@1 metric of TextOnly reached 0.6961 (\emph{SD} =  0.0887), and the Hit@5 reached 0.8994 (\emph{SD} = 0.0864), with a satisfaction rating of 4.26 (\emph{SD} = 0.41). It is evident that TextOnly's recommendation capability exhibits considerable variation among different participants. Notably, the accuracy averaged across participants is lower than the overall acurracy. This is due to the exceptionally low data volume for certain participants, which tends to adversely affect TextOnly's performance.  Considering the quantity of options available and the fact that some inputs are inherently suitable for multiple functions, the current Hit@1 accuracy of TextOnly is considered satisfactory. However, there is a need for further improvement in its Hit@5 accuracy.

\subsubsection{Performance comparison}

Table~\ref{tab:comparison} presents the performance of TextOnly and the compared models. All metrics are averaged across participants, with the standard deviations in parentheses. The first section of the table displays three baselines and their inference time are not listed as the values are too small (on the order of $10^{-5}$). The second section of the table comprises four variants of TextOnly, serving as ablation studies. \textbf{The results clearly demonstrate that TextOnly outperforms other models significantly, with its variants also outperforming the baselines.}

\change{\textbf{The framework of TextOnly is highly effective.}} Among the four variants of TextOnly, TextOnly-nocontext exhibits the best performance, which excludes the use of contextual information. This indicates the overall effectiveness of TextOnly's framework, enabling it to achieve commendable accuracy in the absence of \change{contextual} information. The performances of LLM-only and BERT-only are both \change{significantly inferior compared to TextOnly ($t = -2.985, -3.052$, both $p < 0.01$)}, \change{which combines the LLM and BERT model together}. This demonstrates that LLM and BERT models cooperatively improve each other's inference capabilities within TextOnly's framework.

\change{\textbf{Contextual information contributes to the prediction of user intentions greatly.}} \change{Despite that TextOnly-nocontext performs the best among the variants of TextOnly}, there is still a notable gap between \change{the performances of} TextOnly and TextOnly-nocontext. This indicates the effectiveness of contextual information in this prediction task and suggests potential benefits of incorporating more contextual information. 

\change{\textbf{User personalization is of great significance.}} TextOnly-general demonstrates the lowest accuracy \change{among the four variants of TextOnly}, highlighting that a general model for all users \change{is not suited for this prediction task.} The general model not only fails to provide accurate predictions, but also introduces inappropriate prompts that adversely impact LLM's predictions. \change{This emphasizes the importance and neccesity for maintaining personalized models for each user.}

\change{\textbf{The BERT model can effectively improve the system's inference speed.}} Regarding inference time, it should be noted that we recorded the response time in real usage for TextOnly. For other models, the time listed excludes network latency. Nonetheless, TextOnly still exhibits significantly shorter inference time than LLM-only $(t = -6.905, p < .001)$, indicating that the BERT model gradually takes over the inference task, thereby boosting the model's overall inference speed. The inference time of TextOnly-general is longer compared to TextOnly-nocontext and TextOnly, due to \change{its higher value of threshold $T$, which leads to greater proportion of LLM's predictions.} In this context, its lower accuracy further demonstrates the unsuitability of general models for this \change{prediction task}.

\subsubsection{Performance over time}

\begin{figure}
  \centering
  \subfigure[Accuracy]{
  \begin{minipage}[b]{0.3\linewidth}
    \centering
    \includegraphics[width=\linewidth]{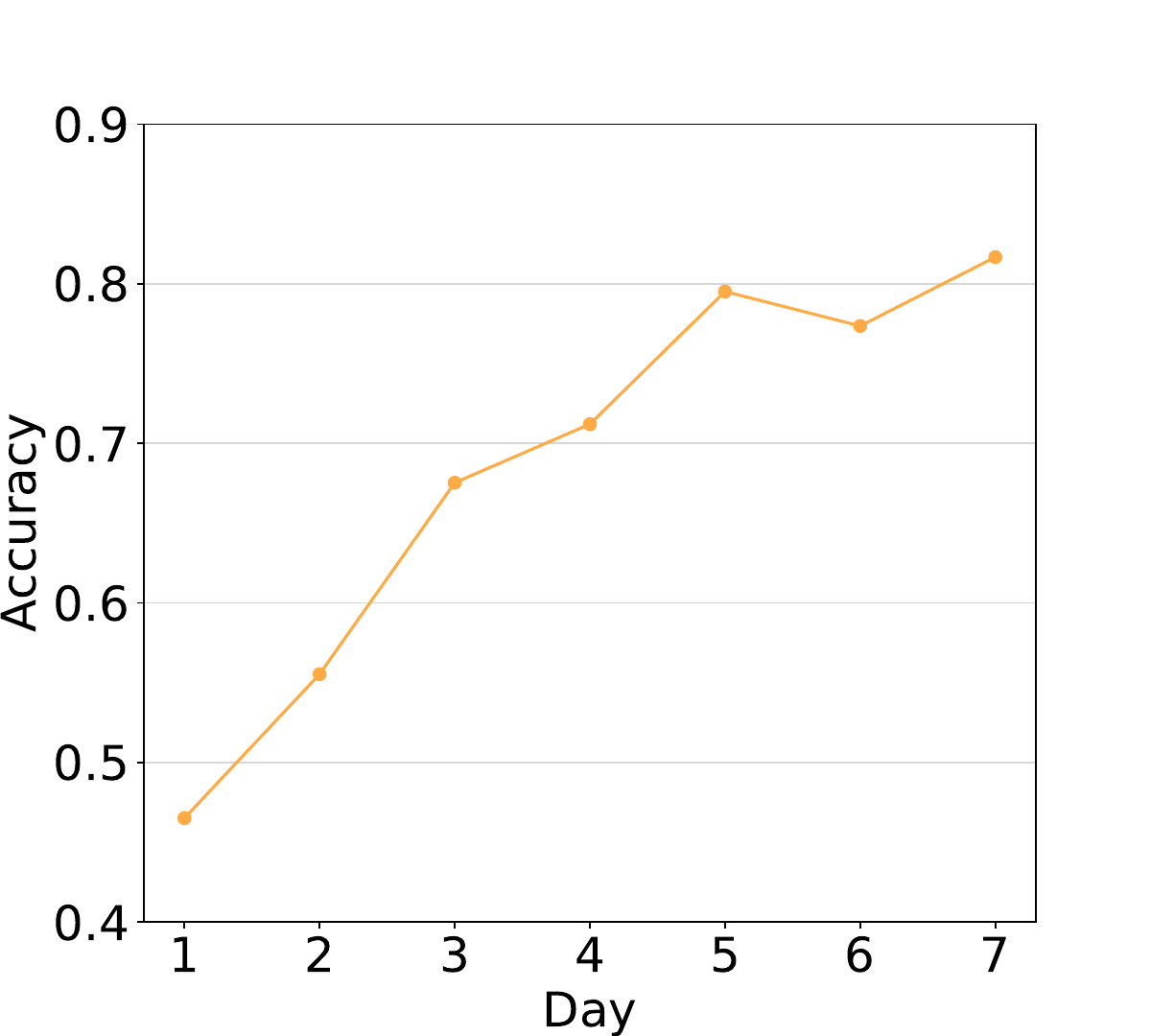}
    \label{fig:acc_by_day}
  \end{minipage}
  }
  \subfigure[Inference time]{
  \begin{minipage}[b]{0.3\linewidth}
    \centering
    \includegraphics[width=\linewidth]{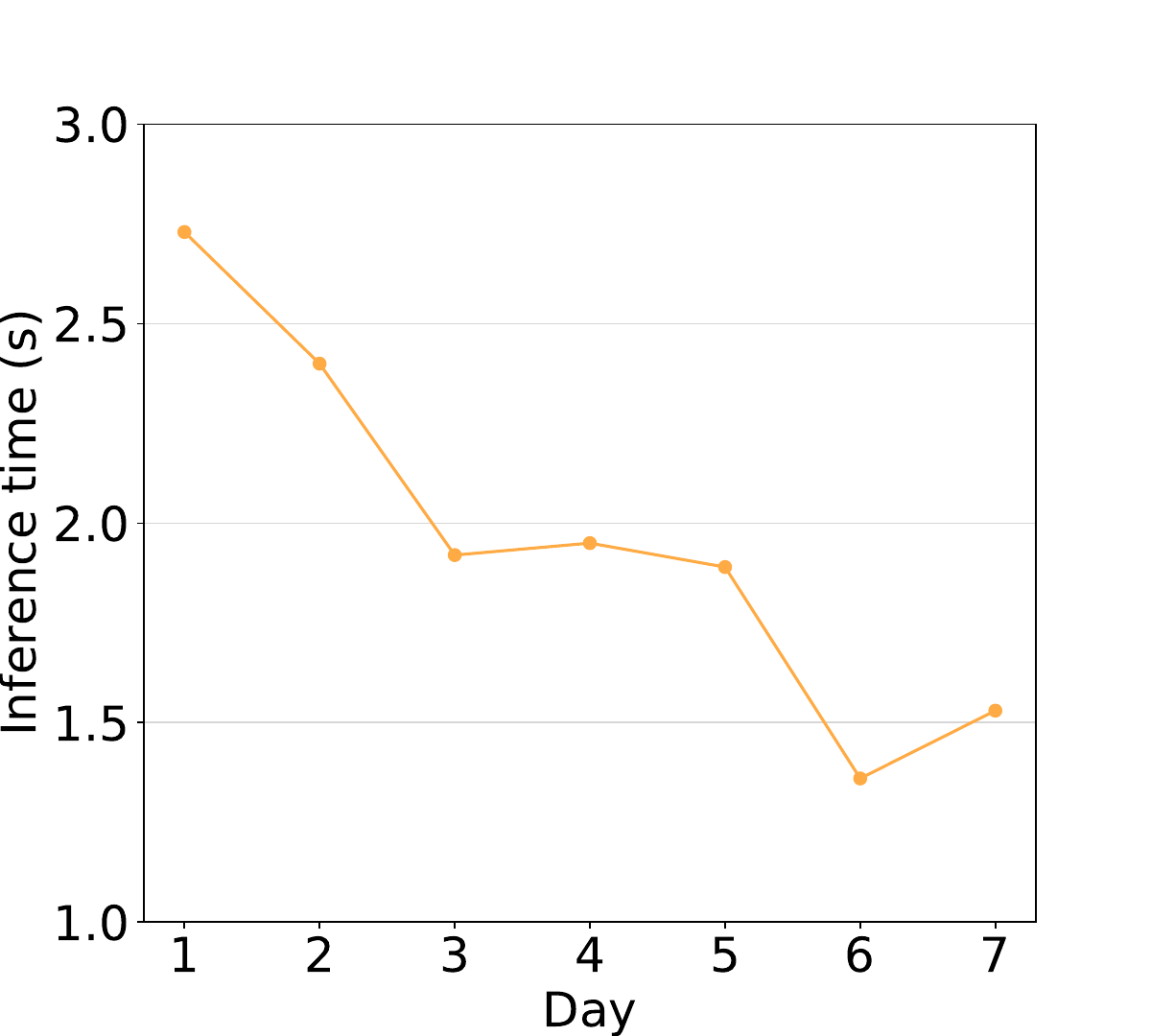}
    \label{fig:time_by_day}
  \end{minipage}
  }
  \subfigure[Number of unique functions]{
  \begin{minipage}[b]{0.3\linewidth}
    \centering
    \includegraphics[width=\linewidth]{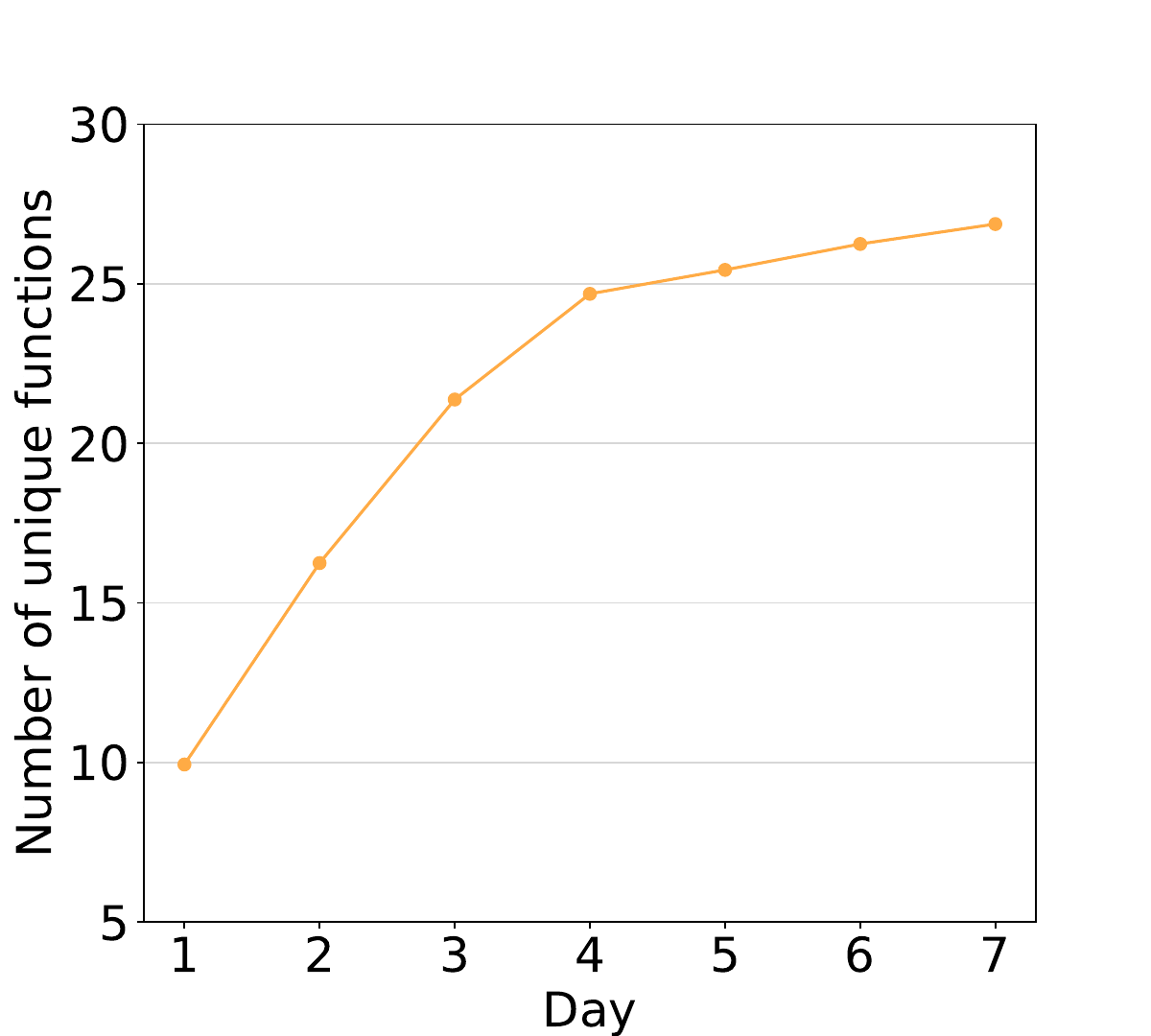}
    \label{fig:intent_by_day}
  \end{minipage}
  }
  \caption{Daily performance of TextOnly during user study}
  \label{fig:stat_by_day}
\end{figure}

As mentioned in Section~\ref{framework}, \change{we incorporated a BERT model to learn from user data and LLM's predictions, to enhance TextOnly's accuracy and inference speed during usage.} \change{To verify this, we tracked TextOnly's performance in each day of the user study.} Figure~\ref{fig:stat_by_day} depicts TextOnly's accuracy, inference time and the accumulated number of used text-related functions on a daily basis. In this context, ``accuracy'' refers to the Hit@1 accuracy.

\change{\textbf{The accuracy of TextOnly significantly improved over time.}} \change{As Figure~\ref{fig:acc_by_day} shows,} TextOnly's accuracy exhibited a significant \change{improvement} throughout the user study, \change{especially for the first few days.} The accuracy on the first day was the lowest, at merely 46.51\%. However, there was a substantial \change{increase in accuracy for} each subsequent day, reaching \change{an impressive accuracy of} 81.68\% on the final day. This trend suggests that TextOnly can effectively learn user-specific knowledge and representations through data accumulation, thus improving its \change{predictive abilities}. \change{On the other hand}, leveraging the capabilities of the LLM, TextOnly is able to achieve a moderate accuracy at the very beginning of its use, \change{addressing the cold start problem effectively.}

\change{\textbf{The inference speed of TextOnly also improved remarkably during usage.}} Figure~\ref{fig:time_by_day} reveals that \change{TextOnly's inference time} noticeably decreased as the user study progressed. \change{It dropped from 2.73 seconds on day one to 1.36 seconds on day six.} The BERT model took over 72.84\% of predictions on day six and 67.66\% on day seven. For comparison, the BERT model took only 0.18 seconds per prediction, \change{while} the LLM took 4.32 seconds on average. This confirms our initial hypothesis that TextOnly will \change{gradually} decrease its reliance on the LLM for predictions with continued usage, thereby improving system's overall inference speed.

Figure~\ref{fig:intent_by_day} depicts the cumulative number of unique text-related functions used by participants \change{during the} user study. The number increased swiftly in the early days and then leveled off in the latter days. \change{This trends indicates that the collection of participants' text-related functions began to stabilize.} Furthermore, it is important to note that the growth of candidate functions did not adversely affect the improvement in TextOnly's accuracy and inference speed.

In summary, the results of our real-world user study indicate that \textbf{TextOnly is capable of achieving high accuracy and rapid inference speed after an initial usage of 3-4 days.} This convincingly validates the practicality of TextOnly.

\subsubsection{\change{Participant feedback}}

\change{
Following the user study, we conducted semi-structured interviews with the participants. We first posed some preset questions to the participants, followed by open discussions based on their responses. We presented the following results.}

\change{
\textbf{The accuracy of TextOnly is fairly satisfactory.} Out of the 16 participants, 13 expressed satisfaction with the recommendation accuracy of TextOnly, or found it exceeded their expectations. P15 mentioned that \emph{``its recommendations are quite accurate, and I can feel it adapting to my habits''}. P8 suggested that TextOnly's performance exceeded his expectations, as he thought his intentions were hard to predict sometimes, yet TextOnly's predictions turn out to be accurate. For participants who are not very satisfied, P5 stated that the recommendations of TextOnly were not accurate enough, as \emph{``I often use multiple apps to search for the same content, and this seems to confuse the system''}. P3 found TextOnly's recommendations not accurate at the beginning of use, but gradually improves over time. Similarly, most participants perceived an evident improvement in TextOnly's recommendation capabilities during usage.}

\change{
\textbf{Participants prefer TextOnly over manual execution.} Compared to performing tasks by themselves, participants expressed a preference for using TextOnly for its efficiency. Participants stated that in specific scenarios, they particularly prefer TextOnly, such as \emph{``in non-urgent circumstances (P1, P6, P8 and P10)''}, \emph{``for commonly used functions (P3, P7, P14, P16)''}, \emph{``when the phone is locked or on the home screen (P4, P10)''}, and \emph{``when the desired application is not on the current screen (P2, P11)''}. Besides, P5, P7 and P14 also mentioned that sometimes they are not sure which app is more suitable for their intentions, and TextOnly appears to be a good choice in such cases.}

\change{
\textbf{The inference time of TextOnly is acceptable, but better to be shorter.} Of the 16 participants, 12 found the waiting time acceptable or even short, while nearly half of them also expressed hope that it could be shorter. Four other participants (P2, P3, P7, P10) found the waiting time to be relatively long. In particular, P7 stated that the large variance in waiting time negatively impacted her use experience. This issue arises from the significant difference in inference time between the BERT model and LLM, and needs to be addressed in future work.}

\change{
\textbf{Participants have good patience and confidence in TextOnly, even when it is inaccurate.} Most participants expressed willingness to invest time to ``train'' TextOnly for its better performance in future use, except for P5 and P13, who stated that inaccurate results affected their confidence in future use. When TextOnly's recommendations were not accurate enough, some participants (P1, P9, P11, P16) expressed understanding, acknowledging the inherent difficulty of the task. P6, P12 and P14 expressed confidence that given this trial, TextOnly would be capable of handling similar situations next time. P7 and P8 also mentioned that although sometimes the recommended function was not exactly what they expected, it could still meet their intentions well.}

\change{
\textbf{TextOnly can change users' input habits to some extent.} Through usage, most participants have gained a certain understanding of TextOnly's capabilities, recognizing that inputs with evident characteristics or format are more likely to yield satisfactory results. This, to some extent, could change their input behaviors. For instance, P4 stated that she would \emph{``simplify the keywords for searching to distinguish them from chatting inputs, like typing `cold medication' instead of `What medicine should I take for a cold?' ''}. P11, P14, P15 also mentioned that they would deliberately differentiate their input under different intentions. On the other hand, participants won't make substantial modifications to their input, as it will be directly entered into their desired text boxes.}

\subsection{Laboratory \change{User} Study\label{labstudy}}

\change{In the real-world user study, we excluded participants' chat-oriented intentions due to privacy concerns, and the participants only used TextOnly without comparing it with other methods. Therefore,} we further conducted a controlled laboratory user study, to evaluate TextOnly's performance on chat-oriented intentions, and compare it with off-the-shelf text portals and voice assistants.

\subsubsection*{\change{Participants}}

We recruited 12 participants (3 males and 9 females), \change{aged between 19 and 26 years old. Detailed information of participants and their smartphones is listed in Appendix~\ref{appendix:participants3}.} The participants were invited to our laboratory to conduct the study. \change{This study is IRB-approved and participants all signed consent forms. Participants received compensation according to the time of the user study (100 CNY per hour).}

\subsubsection*{Procedure}

The study comprised two phases. \change{In the first phase, we} assessed TextOnly's performance for chat-oriented intentions. To simplify, we limited the chat app to Wechat, as it encompasses a dominant proportion of chatting activities for Chinese users. Each participant was required to voluntarily upload chat histories in previous two weeks with ten different contacts or chat groups. All contacts and chat messages were selected by the participants, and they were allowed to skip any messages they didn't want to upload. Participants were paid based on the volume of the chat messages they uploaded. Subsequently, each participant was asked to conduct 10 chat trials using TextOnly to assess its ability to recommend appropriate contacts. 

In the second phase, we compared TextOnly with some off-the-shelf text portals and voice assistants. Participants were instructed to first accomplish certain intentions using TextOnly, and then repeated same intentions with the built-in text portals and voice assistants on their smartphones. The details are as follows:
\begin{itemize}
    \item Built-in text portal: For each test case of TextOnly, the same text input was fed to this portal and we recorded whether it was capable of each corresponding intents. If successful, participants then rated a satisfaction score on a scale of 1-5 for the results provided.
    \item Built-in voice assistant: For each test case of TextOnly, the participants were asked to issue a natural language command describing the same intent. The input command and whether the voice assistant was capable of each corresponding intent were recorded.
\end{itemize}
The specific information of the device used by each participant is presented in Appendix \change{\ref{appendix:participants3}}.

\subsubsection* {Results}

\begin{table}
  \caption{Comparison with the built-in text portal entry and voice assistant}
  \label{tab:comparison}
  \begin{tabular}{llll}
    \toprule
    Model & \% Capable functions & Satisfaction score & length of input \\
    \midrule
    TextOnly & 100.00\% & 4.20 (1.24) & 5.20 (2.95) \\
    Built-in text portal & 52.59\% & 2.86 (1.55) & 5.20 (2.95) \\
    Built-in voice assistant & 66.23\% & - & 11.69 (4.07) \\
  \bottomrule
\end{tabular}
\end{table}

\begin{figure}
  \centering
      \includegraphics[width=1\linewidth]{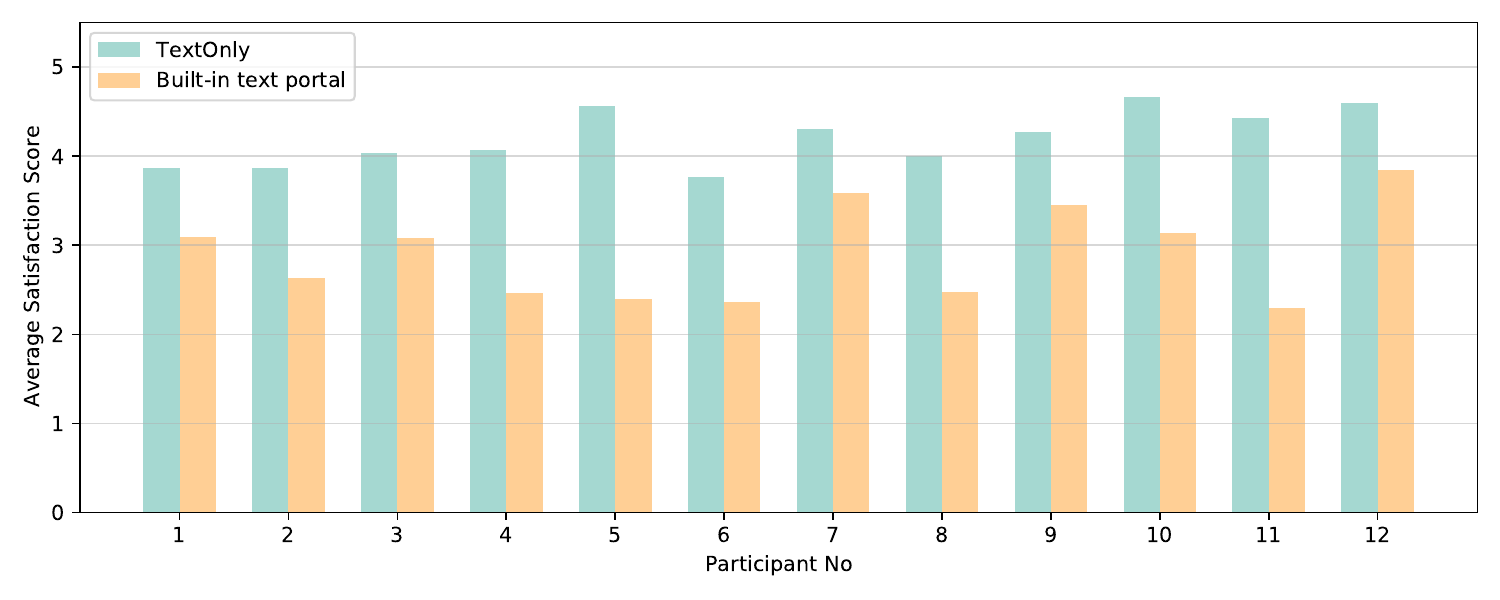}
      \caption{Comparison of satisfaction score between TextOnly and built-in text portal}
      \Description{}
      \label{fig:satisfaction_score_comparison}
\end{figure}

\change{\textbf{TextOnly is capable of identifying chat-oriented intentions and predicting chat contacts.}} For chat-oriented intentions, a total of 120 tests were conducted. TextOnly achieved an accuracy of 96.01\% in identifying \change{chat-oriented intentions}. In \change{the} subsequent task of contact prediction, TextOnly achieved a Hit@1 accuracy of 0.4833, a Hit@5 accuracy of 0.8000, and an MRR of 0.6266. This accuracy is not very high, primarily due to the lack of \change{personalized} training. \change{Moreover, in cases of new topics, the correlation between the new message and previous messages can be quite weak, which increases the difficulty of contact prediction.} However, the accuracy is consistent with that \change{of the first day in the real-world user study. Therefore, it's reasonable to expect that the accuracy of contact prediction will also improve considerably with the accumulation of user data.}

\change{\textbf{TextOnly supports more functions and has higher satisfaction scores than the built-in text portal.}} Table~\ref{tab:comparison} shows the comparison between TextOnly and the two baselines, built-in text portal and voice assistant. \change{Due to the lack of a comprehensive set of intentions} for evaluating these three tools, \change{this} study is confined to \change{evaluate how many} of the TextOnly's achievable intentions can be realized by the two baselines. As a result, the built-in text portal accomplished 52.59\% of the functions, \change{slightly more than} a half of the test set. We find it mainly capable of searching tasks and a small number of default services, such as taking notes and retrieving contacts. For its capable functions, \change{its} satisfaction score is significantly lower than that of TextOnly $(W = 3508.00, p < .001)$ according to the Wilcoxon signed-rank test, which is also reflected in Figure~\ref{fig:satisfaction_score_comparison}. 

\change{\textbf{TextOnly supports more functions and shorter inputs than the built-in voice assistant.}} The built-in voice assistant supports a wider range of functions than the text portal. However, due to the limited integration with third-party applications, it can only perform basic tasks and is not as flexible as TextOnly in terms of function customization. Furthermore, the voice assistant necessitates more extensive inputs to comprehensively articulate \change{users'} intended function. From Table\ref{tab:comparison}, we can observe that the average input length of voice assistant is more than twice as long as TextOnly's. While voice input is more convenient than typing, it is important to note that  TextOnly does not limit its input modality to typing in either its conception or implementation, and is capable of voice input as well. Consequently, the significantly shorter input is indeed an advantage of TextOnly in comparison to voice assistants. 

\subsection{Discussions}

Based on the results of the \change{above} user studies, we conclude that TextOnly demonstrated commendable accuracy in practical use, and its \change{accuracy and inference speed} continuously improved during usage. \change{Participants are fairly satisfied with TextOnly's performance, and prefer to use TextOnly over manual executions.} Compared to \change{existing} built-in text portals and voice assistants on smartphones, TextOnly supports a greater range of functions and allows for more concise inputs, respectively. Ablation studies also verified the effectiveness of each module within TextOnly.

\change{Through user studies, we discovered some factors that affect TextOnly's accuracy negatively. On the one hand, some text input may correspond to multiple distinct intentions.} For example, the phrase ``human-computer interaction'' could be a search keyword in various search engines, a short message to be sent in chatting apps, a topic to be recorded in memo, or an item to be filled in a questionnaire, etc. We employed contextual information to distinguish these similar intentions, \change{but in some cases}, recent app usage and notifications \change{do not sufficiently reflect} the user's actual intention. On the other hand, we observed a tendency among users to use various search engines for identical keywords. Sometimes, users conduct searches on multiple search engines concurrently in order to gather additional information. In some cases, however, they \change{used} one application today and switched to another tomorrow for \change{similar or identical inputs}. Consequently, under certain circumstances, users' historical data might inadvertently misguide the predictions.

As an innovative interaction method, TextOnly not only simplifies the interactions of text-related functions, but also extends the scope of NLIs to encompass novel usage scenarios. For instance, via TextOnly, users can trigger the scanning function and complete a payment by simply entering a number. This particular scenario falls outside the conventional realm of text-related functions, and users may not consider using traditional NLIs for such purposes. However, with TextOnly, it can be quickly accomplished through text input. Similarly, users can easily set an alarm by entering a specific time, or access a friend's recent social media posts by \change{simply} typing their name in TextOnly. These examples demonstrate how TextOnly empowers people to accomplish tasks using natural language in a entirely novel way.

\change{In previous research, Aliannejadi et al.\cite{aliannejadi2018target} proposed a unified search framework that recommends search apps based on user queries, and achieved a Hit@1 accuracy of 62.85\% across 121 apps. However, in their dataset, the top 17 apps accounted for over 95\% of the queries, and they utilized 70\% of the data for training. In terms of next-app prediction, Chen et al.\cite{chen2019cap} achieved an 84\% Hit@5 accuracy using Graph-based embeddings, and Yu et al.\cite{yu2018smartphone} reached an 83\% Hit@5 accuracy using transfer learning. In these studies, user data needs to be collected for model training, and the model's performance could be significantly influenced by the quality of data and its alignment with test data. Khaokaew et al.\cite{khaokaew2024maple} leveraged large language models for next-app prediction, achieving a Hit@1 accuracy of 71.57\% on the LSapp dataset\cite{aliannejadi2021context}. In comparison, TextOnly does not require data for pre-training and achieves comparable accuracy at a finer granularity, i.e. function level. In terms of evaluation, TextOnly proved its practicality in real-world user study, rather than offline testing on pre-collected datasets. Most importantly, TextOnly introduces a new interaction method, and help users accomplish more specific intentions.}

\section{LIMITATIONS AND FUTURE WORK}

\begin{figure}
  \centering
  \includegraphics[width=0.8\linewidth]{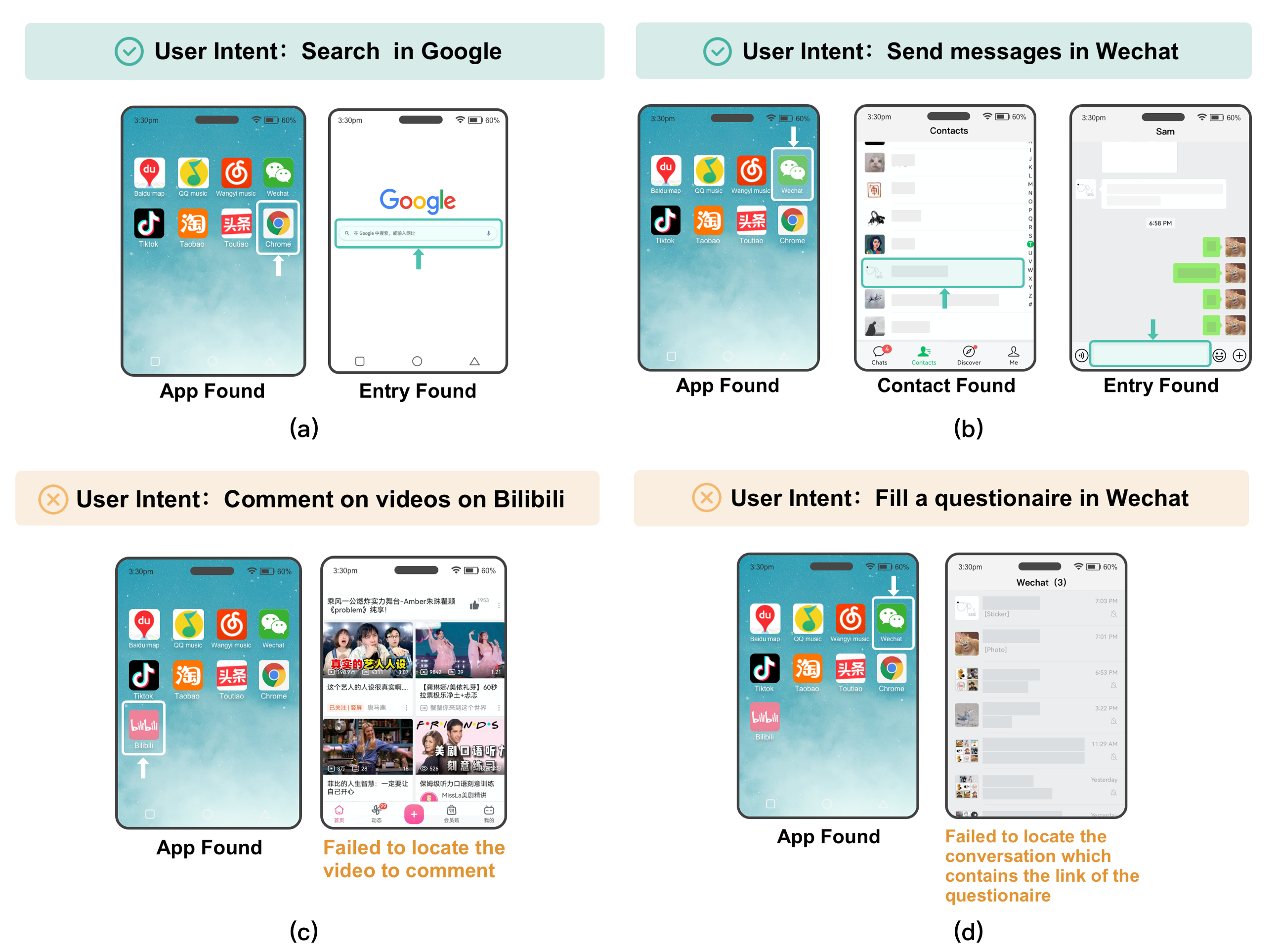}
  \caption{Examples of supported functions and unsupported functions of TextOnly. The text box in (a) is fixed and the text box in (b) can be inferred from chat histories. The text boxes in (c) and (d) are unable to locate as TextOnly lacks enough information.}
  \Description{}
  \label{fig:examples}
\end{figure}


Our current work has preliminarily realized a unified function portal to facilitate quick access to text-related functions, but it still exhibits certain limitations and requires further enhancement. Most of all, the time consumption of our current model is \change{not low enough, negatively affecting users' experience}. This is primarily due to our approach of utilizing LLM through HTTP requests, which adds extra time for network transmission. To alleviate this issue, alternative offline or lightweight versions of LLM can be employed, \change{which can significantly reduce prediction time. }

At present, TextOnly does not support all text-related functions. Certain text-related functions require more parameters than only text inputs, making it challenging for TextOnly to differentiate and target these functions. Figure~\ref{fig:examples} illustrates some use cases that TextOnly supports and does not support. For those unsupported cases, TextOnly struggles to locate the specific text box even though it has identified the correct app and action, because the desired functions need more parameters to be identified. The chat-oriented intentions also include contacts as an additional parameter, so we had to make specific adjustments to accommodate them in TextOnly. To accommodate these currently unsupported functions, we consider capturing the text boxes that appears in the recent interfaces for a dynamic set of candidate functions. This approach not only prevents an excessive number of candidates for prediction, but also adequately covers potential use cases.

In terms of usage, users currently has to record new RPA scripts in order to incorporate new text-related functions, which \change{can be} a complex and learning-intensive task. To mitigate this, a platform can be established for users to share RPA scripts, thereby offering broader selections to users and reducing the need for recording RPA scripts on their own. However, to truly address this issue, we aim to automate the recording of new functions, which is complementary to our intention to capture text boxes appearing in user interfaces.

\change{Intriguingly, with the adoption of LLM as the main inference module, our framework capitalizes on LLM's extensive knowledge and capabilities, yet it inherently faces a ceiling in accuracy dictated by the LLM's limitations. It is foreseeable that large language models will continue to undergo further development and improvement in the late future. In this case, with the increasing capabilities of LLM, TextOnly is expected to exhibit better performance, even without modifications on current framework.}

\section{CONCLUSIONS}

In this paper, we present TextOnly, a unified function portal which offers quick access to the text-related functions on smartphones by users' raw text inputs. \change{By reducing the need to describe users' intentions in detail}, TextOnly effectively improves \change{interaction efficiency} compared to voice assistants. We integrated a pre-trained LLM for general knowledge to address the cold start problem \change{on new users and new functions}. We also incorporated a BERT model that learns from user data and LLM’s predictions, which reduces time consumption and economic costs for predictions.

\change{In} a week-long real-world user study with 16 participants, we demonstrated the practicality of TextOnly, which achieved an accuracy of 71.35\% for top-1 recommendations and 89.94\% for top-5 recommendations. During the study, TextOnly exhibited significant improvements over time in both accuracy and inference speed, demonstrating its capacity for user personalization. \change{The participants are fairly satisfied with TextOnly's performance and have a preference for TextOnly over manual executions.} \change{Through} ablation studies, we \change{validated} the effectiveness of our framework, \change{where} LLM and the BERT model cooperate and enhance each other's performance. \change{Furthermore, a laboratory study} showed that TextOnly outperforms built-in text portals and voice assistants on smartphones in terms of the number of supported functions and \change{the length of inputs}.

Our work proposes a novel interaction approach for text-related functions on smartphones and broadens the scope of natural language interfaces by extending the boundaries of usage patterns and supported tasks. We also provide an example of integrating large language models with traditional models to address practical human-computer interaction problems.

\bibliographystyle{ACM-Reference-Format}
\bibliography{sample-base}

\clearpage
\appendix

\section{Detailed information of participants}

\subsection{Participants in Data Collection\label{appendix:participants1}}

\begin{table}[h]
  \caption{Detailed information of participants in data collection}
  \label{tab:appendix-participants1}
  \begin{tabular}{ccc|ccc}
    \toprule
    Participant No   & Age   & Gender &  Participant No   & Age   & Gender \\
    \midrule
        P1 & 22 & F & P12 & 24 & M \\
        P2 & 25 & M & P13 & 23 & F \\
        P3 & 21 & F & P14 & 26 & F \\
        P4 & 23 & F & P15 & 22 & M \\
        P5 & 19 & F & P16 & 22 & F \\
        P6 & 20 & M & P17 & 23 & M \\
        P7 & 22 & M & P18 & 24 & F \\
        P8 & 21 & F & P19 & 20 & F \\
        P9 & 22 & F & P20 & 30 & M \\
        P10 & 22 & M & P21 & 19 & F \\
        P11 & 24 & F & P22 & 23 & F \\
  \bottomrule
\end{tabular}
\end{table}

\subsection{Participants in User Study\label{appendix:participants2}}

\begin{table}[h]
  \centering
  \caption{Detailed information of participants in user study}
  \label{tab:appendix-participants2}
  \begin{tabular}{cccc}
    \toprule
    Participant No   & Age   & Gender & Phone Model \\
    \midrule
    P1 & 23 & M & Xiaomi 10 \\
    P2 & 22 & F & Vivo s9 \\
    P3 & 23 & F & Huawei nova9 \\
    P4 & 19 & F & Vivo x80 \\
    P5 & 22 & F & Vivo iQoo neo6 \\
    P6 & 22 & M & Meizu 20 pro \\
    P7 & 28 & F & Redmi note12 pro \\
    P8 & 24 & M & Redmi k50 \\
    P9 & 26 & F & Huawei nova7 \\
    P10 & 26 & M & Redmi k30 pro \\
    P11 & 24 & F & Yijia 9R \\
    P12 & 23 & M & Vivo x90 pro \\
    P13 & 22 & F & Xiaomi 12s \\
    P14 & 22 & F & Xiaomi 11 lite \\
    P15 & 22 & M & Yijia ace2 \\
    P16 & 22 & F & Honor magic5 \\
  \bottomrule
\end{tabular}
\end{table}

\clearpage
\subsection{Participants in Laboratory Study\label{appendix:participants3}}

\begin{table}[h]
  \centering
  \caption{Detailed information of participants in laboratory study}
  \label{tab:appendix-participants3}
  \begin{tabular}{cccc}
    \toprule
    Participant No   & Age   & Gender & Phone Model \\
    \midrule
    P1 & 20 & F & Huawei P30 \\
    P2 & 22 & F & Xiaomi 10s \\
    P3 & 19 & F & Xiaomi cc9 pro \\
    P4 & 23 & M & Huawei nova10 \\
    P5 & 26 & F & Huawei nova5 pro \\
    P6 & 22 & F & Vivo x80 \\
    P7 & 22 & F & Huawei nova10\\
    P8 & 24 & M & Vivo iQoo9\\
    P9 & 21 & M & Honor magic4\\
    P10 & 19 & F & Huawei mate20\\
    P11 & 23 & F & Redmi k40\\
    P12 & 21 & F & Oppo reno5\\
  \bottomrule
\end{tabular}
\end{table}

\clearpage
\section{Prompt for LLM\label{appendix:prompt}}

\subsection{Prompt for function prediction\label{appendix:prompt1}}

\begin{figure}[h]
  \centering
  \includegraphics[width=1.0\linewidth]{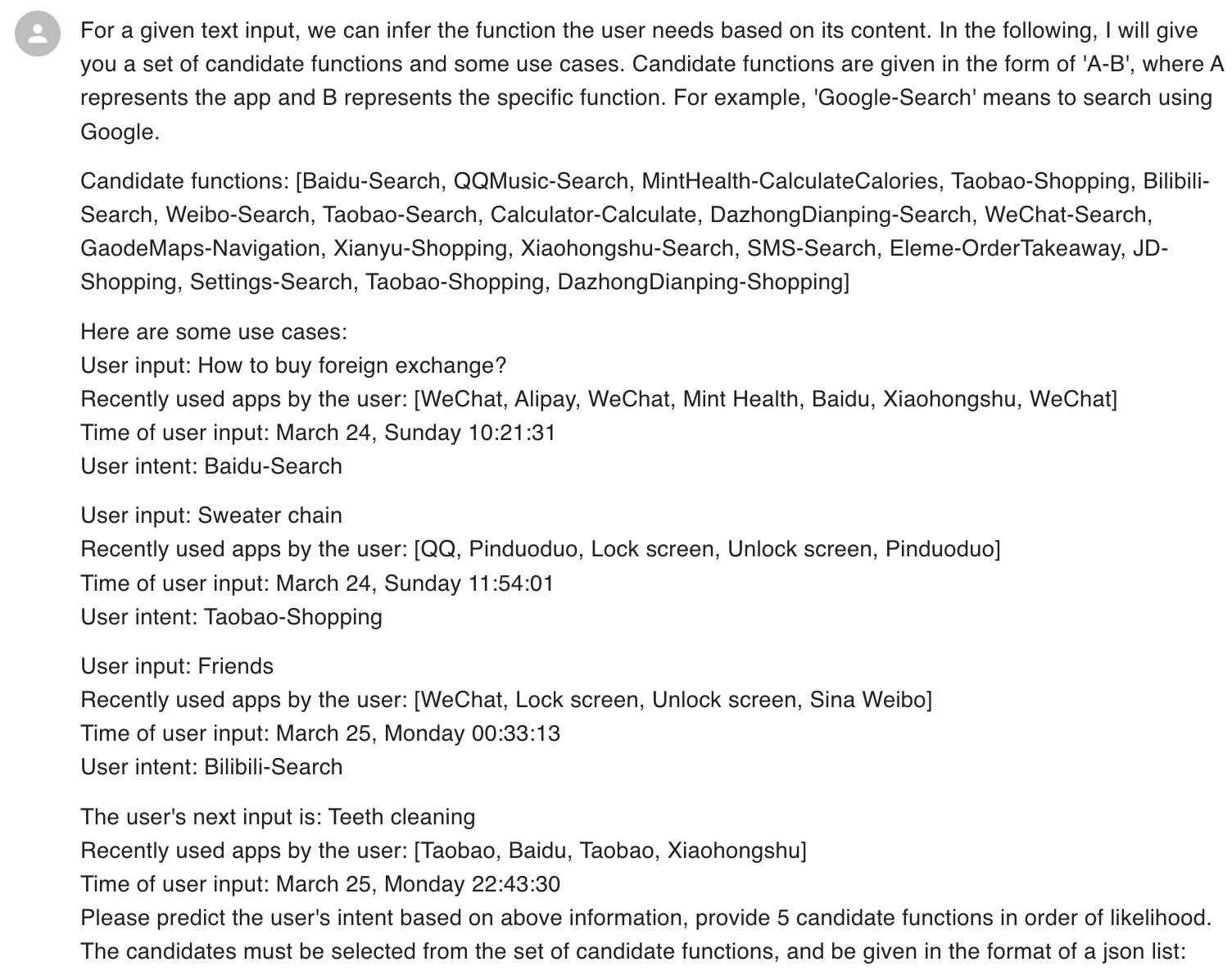}
  \caption{An example of the prompt for function prediction. In real usage, there are more use cases.}
  \label{fig:prompt}
\end{figure}

\clearpage
\subsection{Prompt for contact selection\label{appendix:prompt2}}

\begin{figure}[h]
  \centering
  \includegraphics[width=1.0\linewidth]{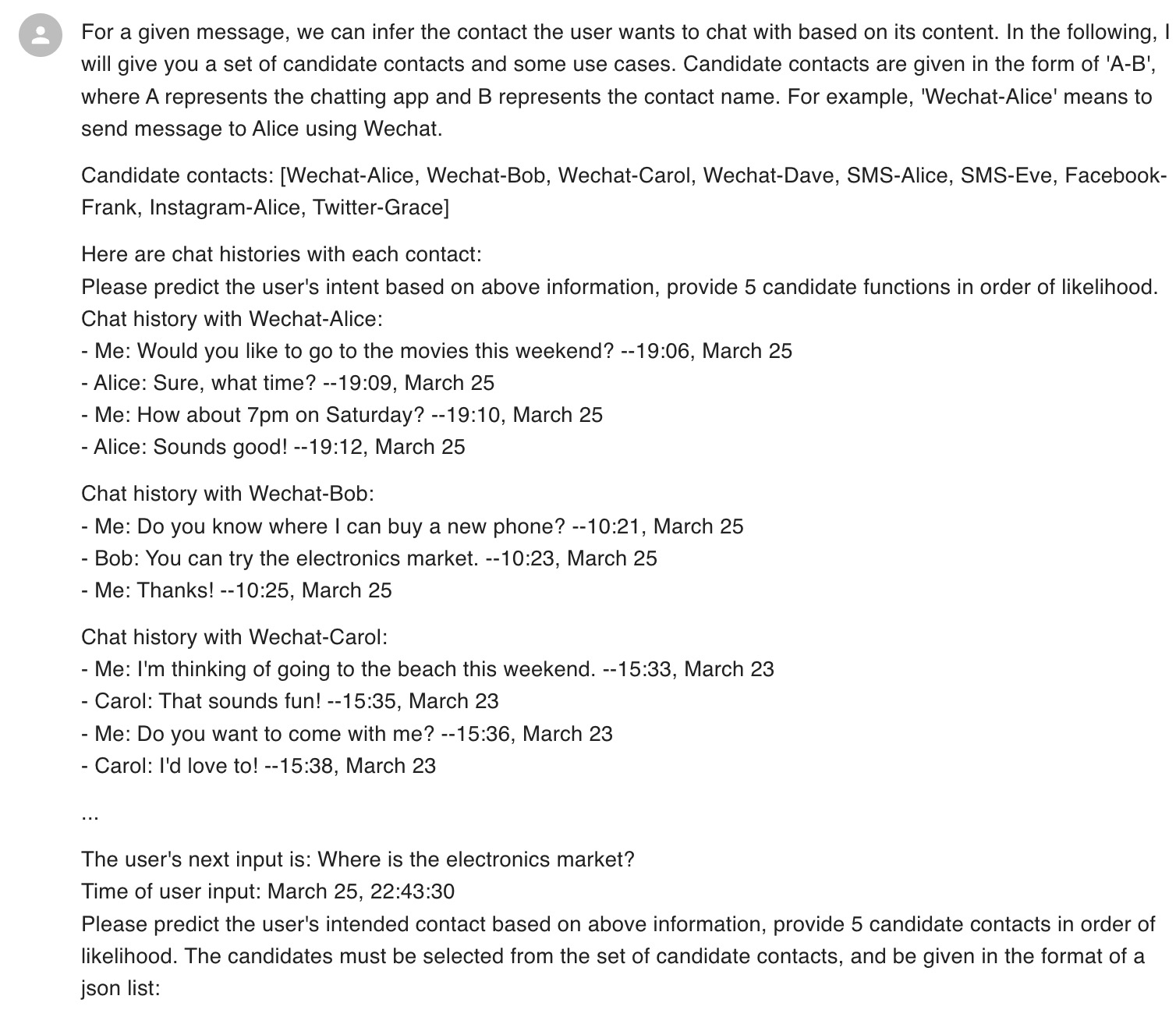}
  \caption{An example of the prompt for contact selection. In real usage, chat histories with all contacts are included.}
  \label{fig:prompt2}
\end{figure}







\end{document}